# BowNet: Dilated Convolution Neural Network for Ultrasound Tongue Contour Extraction

*M. Hamed Mozaffari*[1*], *Won-Sook Lee*[1]

[1]*School of Electrical Engineering and Computer Science, University of Ottawa, 800 King-Edward Avenue, Ottawa, Ontario, Canada, K1N-6N5.*
[*]*mmoza102@uottawa.ca*

**Abstract**

Ultrasound imaging is safe, relatively affordable, and capable of real-time performance. This technology has been used for real-time visualization and analyzing the functionality of human organs in many studies. One application of this technology is to visualize and to characterize human tongue shape and motion during a real-time speech to study healthy or impaired speech production.

Due to the noisy nature of ultrasound images with low-contrast characteristic, it might require expertise for non-expert users to recognize organ shape such as tongue surface (dorsum). To alleviate this difficulty for quantitative analysis of tongue shape and motion, tongue surface can be extracted, tracked, and visualized instead of the whole tongue region. Delineating the tongue surface from each frame is a cumbersome, subjective, and error-prone task. Furthermore, the rapidity and complexity of tongue gestures have made it a challenging task, and manual segmentation is not a feasible solution for real-time applications.

The progress of deep convolutional neural networks has been successfully exploited in various computer vision tasks such as image classification and segmentation. Several end-to-end deep learning segmentation methods provide a promising alternative for previous techniques with higher accuracy and robustness results, without any intervention. Employing the power of high-speed graphics processing unit (GPU) with state-of-the-art deep neural network models and training techniques, it is feasible to implement new fully-automatic, accurate, and robust segmentation methods with the capability of real-time performance, applicable for tracking of the tongue contours during the speech.

This paper presents two novel deep neural network models named BowNet and wBowNet benefits from the ability of global prediction of decoding-encoding models, with integrated multi-scale contextual information, and capability of full-resolution (local) extraction of dilated convolutions. Experimental results using several ultrasound tongue image datasets revealed that the combination of both localization and globalization searching could improve prediction result significantly. Assessment of BowNet models using both qualitatively and quantitatively studies showed their outstanding achievements in terms of accuracy and robustness in compare with similar techniques.

## 1 Introduction

Typically, during tongue data acquisition, ultrasound probe beneath the user's chin images tongue surface in midsagittal or coronal view in real-time. Mid-sagittal view of the tongue in ultrasound data is usually adapted for illustration of tongue region, as it displays relative backness, height, and the slope of various areas of the tongue. Tongue dorsum can be seen in this view as a thick, long, bright, and continues region due to the tissue-air reflection of ultrasound signal by the air around the tongue (see Figure 1). This thick white bright region is irrelevant, and the tongue surface is the gradient from white to black area at the lower edge (Stone, 2005). Although the tongue contour region can be seen in ultrasound data clearly, there are no hard structure references. This makes it difficult to localize the tongue position and interpret its gestures (Stone, 2005). Furthermore, due to the noise characteristic of ultrasound images with the low-contrast property, it is not an easy task for non-expert users to localize the tongue surface when real-time tracking is required.

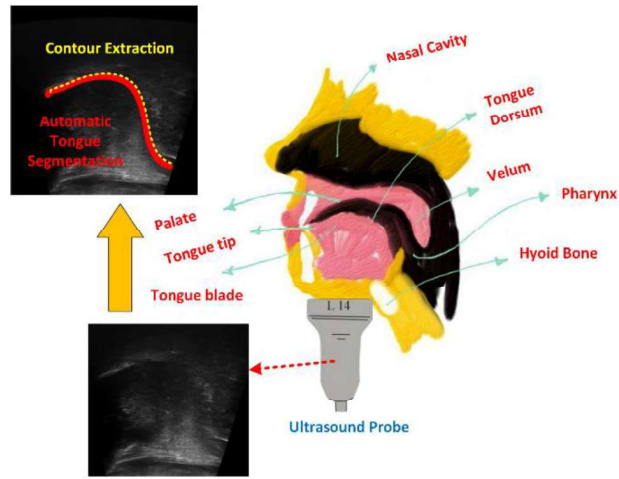

Figure 1. A sample frame of ultrasound tongue imaging data acquired from a mid-sagittal view when the user keeps the ultrasound probe beneath the chin area. Reflection from the tongue dorsum area can be seen as a gradient bright to the black region specified by the red curve. The yellow dotted curve is the tongue contour. On the lower left side of the image, shadow from jaw bone is clearly illustrated.

Ultrasound technology has been a well-known method in speech research for studying of tongue motion and speech articulation (Fasel and Berry, 2010). The popularity of ultrasound imaging for tongue visualization is because of its attractive characteristics such as imaging at a reasonably rapid frame rate which allows researchers to visualize subtle and swift gestures of the tongue during the speech in real-time. Moreover, ultrasound technology is portable, relatively affordable, and clinically safe with a non-invasive nature (Stone, 2005).

For tracking and extracting the tongue contour, researchers have proposed several methods each of which requires some monitoring and modification while the tracking process is in progress. In this study, we introduced a new method for automatic tongue contour extraction in real-time to handle difficulties of previous techniques as we experienced in our research. Main contributions of this paper can be itemized as below:

- Proposing two robust, accurate, fully-automatic deep learning architectures specifically for tongue contour extraction while the proposed models have the capability of real-time performance using GPU power. An alternative solution for CPU only purposes has been reported in previous research (Mozaffari et al., 2018).
- For the first time, dilated convolution is combined with the standard deep, dense classification method in two new network architectures to extract both local and global context from each frame at the same time in an end-to-end fashion. Tongue contour extraction is only an application of these models.
- Creating two comprehensive datasets of tongue images and ground truths, each of which containing two types of data from two different ultrasound machines to test the overfitting problem and robustness of each technique in terms of generalization capability.
- Creating a new customized library in Python for image enhancement and annotation using the B-Spline method to acquire smooth curves in ground truth annotation stage instead of partial straight lines applicable for any ultrasound tongue datasets.
- Enhancing datasets and bringing up the problem of using gray-scale images as the labeled data in previous tongue semantic segmentation studies where 255 class labels have been used instead of 2 classes. Employing the informed undersampling method (Liu et al., 2006) to increase the variety of datasets. Application of skeletonization for achieving tongue contour curves from prediction maps. Evaluation of using offline versus online augmentation in accuracy of segmentation methods where applied on ultrasound tongue data.

We believe that publishing our datasets, annotation package, and our proposed deep learning architectures, all implemented in multiplatform python language with an easy to use documentation can help other researchers in this field to fill the gap of using previous methods where several non-accessible requirements are needed as well as they customized for restricted datasets. The remainder of this paper is structured as follows. A literature review of the field is covered in section 2. Section 3 describes in detail our methodology and architecture of BowNet models. The Experimental results and discussion around our proposed segmentation techniques comprise of system setup and dataset, quantitatively and qualitatively study, and comparison results are presented and analyzed in section 4. Section 5 concludes and outlines future work directions.

## 2 Literature Review and Related Works

Studying and exploiting the dynamic nature of speech data from ultrasound tongue image sequences might provide valuable information, and it is of great interest in many recent studies (Laporte and Ménard, 2018). Ultrasound imaging has been utilized for tongue motion analysis in treatment of speech sound disorders (Eshky et al., 2018), comparing normal and impaired speech production (Laporte and Ménard, 2018), second language training and rehabilitation (Gick et al., 2008), Salient Speech Interfaces (Denby et al., 2010), and swallowing research (Ohkubo and Scobbie, 2018), 3D tongue modeling (S. Chen et al., 2018), to name a few.

The interpretation of different ultrasound images of tongue is a challenging task for non-expert users, and manual analysis of each frame suffers from several drawbacks such as bias depends on skill of the user or quality of the data, fatigue due to the large number of image frames to be analyzed, and lack of achieving reproducible results (Akgul et al., 1999). The classical method for quantification and localization of tongue shape in ultrasound data is to extract the contours of the tongue dorsum using image processing methods (Xu et al., 2016a). Due to the high video capture rate, noisy characteristics of the ultrasound data, different configuration, power, and settings of the ultrasound device, and rapidity of tongue motion, there might be few frames with no valuable or visible information or with non-continues dorsum region. Therefore, it is crucial to have a fully automatic system for the tongue contour extraction. It is even harder for the case of tongue contour tracking in real-time applications (Mozaffari et al., 2018).

Various methods have been utilized for the problem of automatic tongue extraction in the last recent years such as image segmentation like active contour models or snakes (Ghrenassia et al., 2014; Laporte and Ménard, 2015; Li et al., 2005; Xu et al., 2016b, 2016c), graph-based technique (Tang and Hamarneh, 2010), machine learning-based methods (Berry and Fasel, 2011; Fabre et al., 2015; Fasel and Berry, 2010; Jaumard-Hakoun et al., 2015; L. et al., 2012), and many more. A complete list of recent tongue contour extraction techniques can be found in a study by Laporte et al. (Laporte and Ménard, 2018). Although the methods above have been applied successfully on the ultrasound tongue contour extraction, still manual labeling and initialization are frequently needed. Users should manually label at least one frame with a restriction of drawing near to the tongue region (Laporte and Ménard, 2018). For instance, to use Autotrace, EdgeTrak, or TongueTrack software, users should annotate several points on at least one frame.

Deep convolutional neural networks have been the method of choice for many computer vision applications in recent years. They have shown outstanding performance in many image classification tasks as well as object detection, recognition, and tracking (Lin et al., 2017). Dense image classification task might be considered as a segmentation problem when the goal is to categorize every single pixel by a discrete or continuous label (Yu and Koltun, 2015). Depend on the definition of categories, image segmentation might be classified into semantic (delineate desired objects with a label) (Long et al., 2015; Thoma, 2016) or instance (each specific desired object has a unique label) (Li et al., 2015).

With inspiration, modification, and adaptation of several well-known deep classification networks (Krizhevsky et al., 2017; Simonyan and Zisserman, 2014; Szegedy et al., 2014), fully convolutional neural (FCN) networks was successfully exploited for the semantic segmentation problem in a study by (Long et al., 2015). Instead of utilizing a classifier in the last layer, a fully convolutional layer was used in FCN to provide a prediction map as the model's output. From that time, performance of FCN network models have been improved by employing several different operators such as deconvolution (Noh et al., 2015), concatenation from previous layers (Milletari et al., 2016; Ronneberger et al., 2015), using indexed un-pooling (Badrinarayanan et al., 2015), and adding post-processing stages such as CRFs (Chen et al., 2014a).

Many of these innovations significantly improved the accuracy of segmentation results, usually with the expense of more computational costs due to the significant number of network parameters. Furthermore, consecutive pooling layers which are used to improve receptive field and localization invariance (L.-C. C. Chen et al., 2018) in almost all of those methods cause a considerable reduction of feature resolution in decoding section (Chen et al., 2017). To alleviate this issue atrous or dilated convolutions have been proposed and employed recently (Chen et al., 2017, 2014b; L.-C. C. Chen et al., 2018; Hamaguchi et al., 2018; Yu and Koltun, 2015). Dilated convolutions help the network model to predict instances without losing receptive field, without needs of the fully convolutional layer, and with less learnable parameters in comparison to previous FCN methods. Existence of objects at multiple scales is another challenge in semantic segmentation (L.-C. C. Chen et al., 2018) which many recent studies have been focused on solving that issue (Hamaguchi et al., 2018; Lin et al., 2017).

Few studies have been applied deep semantic segmentation for the problem of ultrasound tongue extraction. In (Jaumard-Hakoun et al., 2016), Restricted Boltzmann Machine (RBM) was trained first on samples of ultrasound tongue images and ground truth labels in the form of encoder-decoder networks. Then, the trained decoder part of the RBM was tuned and utilized for prediction of new instances in a translational fashion from the trained network to the test network. To automatically extract tongue contours without any manipulation on a large number of image frames, modified versions of UNet (Ronneberger et al., 2015) have been used recently for ultrasound tongue extraction (Mozaffari et al., 2018; Zhu et al., 2018).

Like adding CRFs as a post-processing stage for acquiring better local delineation (Chen et al., 2014a), more accurate segmentation results can be achieved when multiscale contextual reasoning from successive pooling and subsampling layers (global exploration) is combined with the full-resolution output from dilated convolutions (Yu and Koltun, 2015). In this work, we develop two new deep convolutional neural models (named BowNet and wBowNet) for the problem of ultrasound tongue extraction. The BowNet is a parallel combination of a dense classification architecture inspired by VGG16 (Simonyan and Zisserman, 2014) and UNet (Ronneberger et al., 2015) with a segmentation model inspired by DeepLab network where dilated convolution layers are used without pooling layers (Chen et al., 2017). The wBowNet is the counterpart of the BowNet where a combination of classification and segmentation networks were designed interconnected to support an even higher resolution in prediction outcomes.

As part of this work, we also re-examine the performance of repurposed versions of UNet (Ronneberger et al., 2015) and DeepLab (Chen et al., 2017). We evaluated our proposed architectures on two different datasets of ultrasound tongue images and the experimental results demonstrate that our fully-automatic proposed models are capable of achieving accurate predictions with less learnable parameters and with real-time performance.

## 3 Methodology

### 3.1 Dilated Convolution

The consecutive combination of convolutional and pooling layers in encoding part of a deep network model results in a significantly lower spatial resolution of output feature maps, typically by a factor of 32 for modern current deep learning architectures which is not the desired resolution for the semantic segmentation purposes (Chen et al., 2017). Several ideas have been proposed to reconstruct an input-size segmented output from the course feature map of encoding part. Interpolation (up-sampling) could be the first solution whereas each pixel of the feature map is repeated to increase the image size (Long et al., 2015). This method of down-sampling and up-sampling with inevitable losing information is not beneficial for the task of segmentation when boundary delineation with a high resolution is required. Transposed convolution (sometimes called deconvolution) have been introduced to solve this issue and to recover the low-resolution prediction maps (Zeiler et al., 2010). Transpose convolutional layer operates opposite of a convolution layer where each pixel of the feature map is expanded to the kernel size and superimposed with its neighbor. Although deconvolution improved segmentation result (Noh et al., 2015), it still suffers from checkerboard problem (Odena et al., 2016) as well as increasing the number of learnable parameters. Other techniques such as indexed un-pooling (Badrinarayanan et al., 2015; Zeiler and Fergus, 2014) require memory for saving positions while max-pooling is applied on a previous feature map.

Using dilated convolution (sometimes called atrous convolution) while the dilation factor is increased monotonously through layers, it revealed that the receptive field could be effectively expanded with keeping a spatial resolution. However, the sparsity of dilated kernels does not always cause performance improvement especially for small objects and details (Hamaguchi et al., 2018). To solve this problem, one solution is to decrease the dilation factor throughout the decoding path of the network model similar to the number of kernels in the UNet network (Hamaguchi et al., 2018). Figure 2 illustrates samples of decoding strategies using interpolation, deconvolution, and dilated convolution.

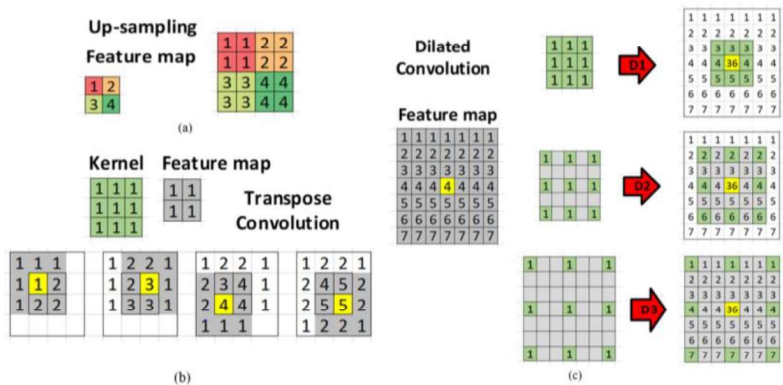

Figure 2. Several ideas for improving the low-resolution feature map to acquire an output image with the input-size image. (a) and (b) Up-sampling using linear interpolation and transpose convolution, respectively. (c) Dilated convolution with different dilation factor of 1, 2, and 3.

## 3.2 Network Architectures

To detect an object such as a tongue contour in relatively noisy ultrasound images, it is vital to pay attention to both context and resolution. In this application, context can be achieved using dense classification methods (Mozaffari et al., 2018). Nevertheless, up-sampling methods such as deconvolution are not able to recover the low-level visual features which are lost in down-sampling stage (Lin et al., 2017), resulting in low-resolution prediction map as the model output. To address this issue, we proposed two deep network models considering both localization and globalization detection.

The BowNet has two separate sub-network forward paths whereas results of the two paths are concatenated in the last layer, followed by a fully convolutional layer. On the contrary, the wBowNet has two interconnected (weaved) sub-networks. Figure 3 and Figure 4 illustrate the BowNet and wBowNet network architecture, respectively. Details of each neural network model used in this study are listed in Table 1. Two popular deep convolutional networks, UNet (Ronneberger et al., 2015) and DeepLab (L.-C. C. Chen et al., 2018; Hamaguchi et al., 2018) are modified and simplified for a fair comparison, and we named them as sUNet and sDeepLab. To have a better criterion for the accuracy of the proposed models, the original UNet was implemented and tested as well.

It can be seen from Table 1 that all the deep convolutional networks can be trained and tested in an end-to-end fashion, which takes an image as input and provides the output as a probability map, directly. In each network architecture, convolutional layers use ReLU activation as a non-linearity function. Drop-out layers and batch normalization layers are followed convolutional layers to improve regularization and accuracy of network models. From Table 1, dilation factor is first increased from one to eight and then decreased again to one in a forward path of successive dilated convolutions to solve the problem of spatial inconsistency (Hamaguchi et al., 2018).

It is common in deep learning semantic segmentation to use SoftMax for the last layer of the network, and at least part of the system is a modified version of VGG16 network (Badrinarayanan et al., 2015; Long et al., 2015; Simonyan and Zisserman, 2014). Moreover, multiclass cross entropy is employed as a loss function for the training of networks. Ultrasound datasets usually contain gray-scale images, and the desired target for tongue contour extraction has only two class labels of tongue contour and background. For this reason, we optimized binary cross-entropy as the loss function, and a fully convolutional layer along with a sigmoid activation function is utilized in the last segment of all the proposed networks. It is noteworthy to mention that receptive filed will be decreased after applying each convolutional layer due to the overlapping of the convolution kernels within the border areas. For this reason, to concatenate feature maps from previous layers, it is necessary to crop feature maps to have similar size images (see faded blue boxes in Figure 3 and Figure 4).

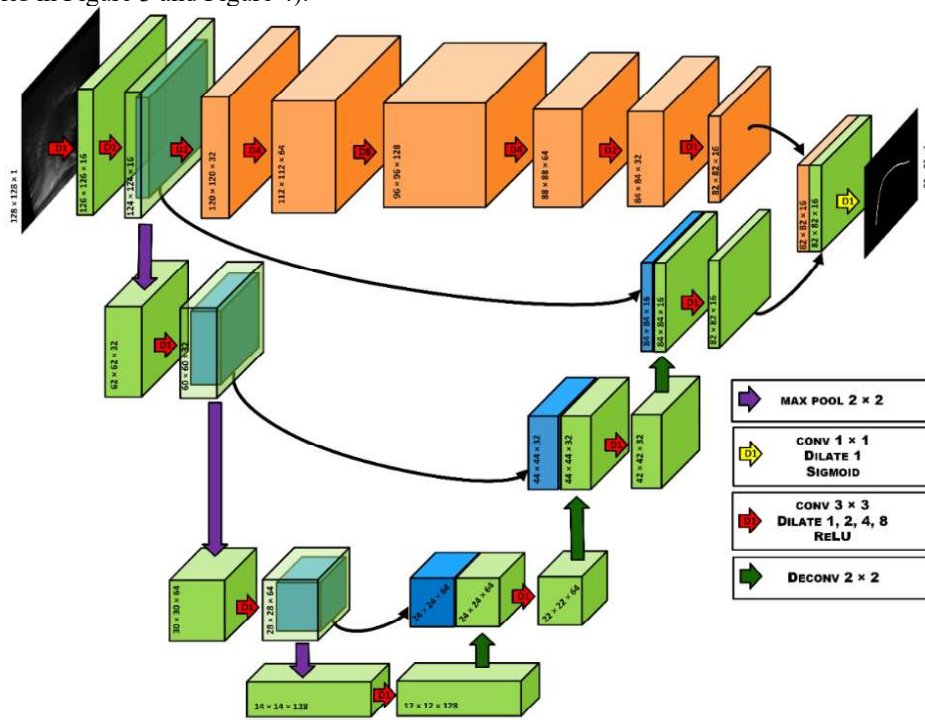

Figure 3. Overview of the proposed BowNet network architecture. In each layer, filter kernels are depicted using boxes. The green and orange boxes are results of standard and dilated convolution layers respectively.

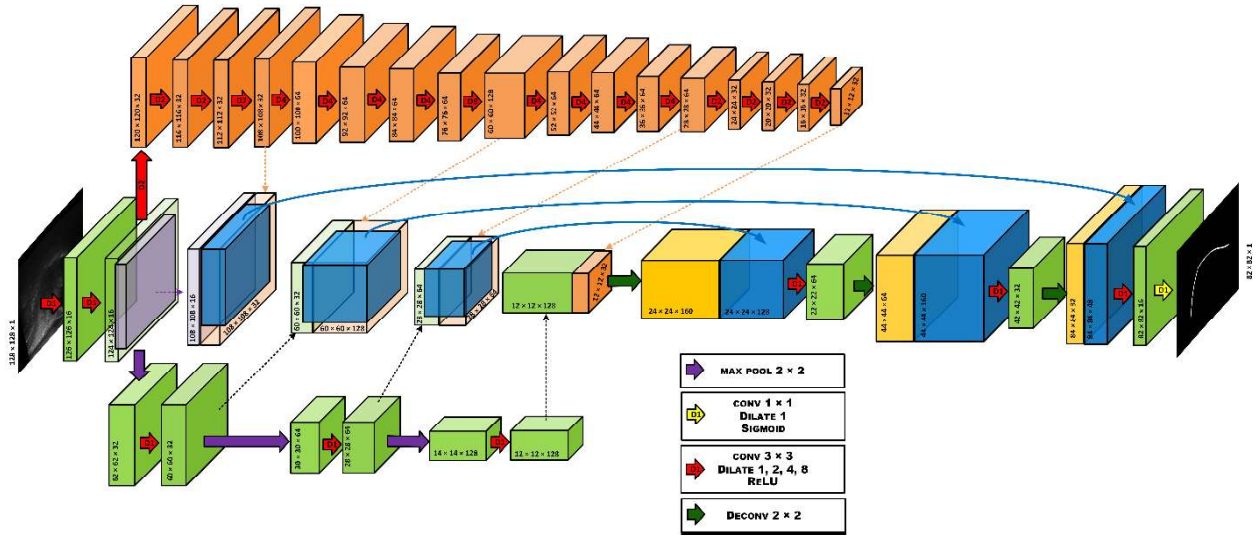

Figure 4. Overview of the proposed wBowNet network architecture. In each layer, filter kernels are depicted using boxes. The green and orange boxes are results of regular and dilated convolution layers respectively.

To fully leverage the power of the trained network for ultrasound tongue contour segmentation, unlike many similar studies with small images as the output (Fasel and Berry, 2010; Jaumard-Hakoun et al., 2015), we keep the image sizes in both input and output layers as 128 × 128 pixels. Input images are cropped and scaled to make them square for the sake of applying convolutional layers, and zero-padding is used to keep the image size throughout the network.

Table 1. The network architecture of BowNet, wBowNet, sUNet, and sDeepLab. In convolution layers, ConvYDX, Y is the number of kernels and X is the value of the dilation factor.

| | NL | wBowNet | | NL | BowNet | | NL | sUNet | NL | sDeepLab |
|---|---|---|---|---|---|---|---|---|---|---|
| INPUT 128 × 128 | 1 | GRAYSCALE | | 1 | GRAYSCALE | | 1 | GRAYSCALE | 1 | GRAYSCALE |
| | | Path-1 | Path-2 | | Path-1 | Path-2 | - | - | - | - |
| CONV-1 | 2 | Conv16 | - | 2 | Conv16 | - | 2 | Conv16 | 2 | Conv16 |
| POOL-1 | 1 | MaxPool | - | 1 | MaxPool | - | 1 | MaxPool | - | - |
| CONV-D2 | 4 | - | Conv32D2 | 1 | - | Conv32D2 | - | - | 1 | Conv32D2 |
| CONV-2 | 1 | Conv32 | - | 1 | Conv32 | - | 1 | Conv32 | - | - |
| POOL-2 | 1 | MaxPool | - | 1 | MaxPool | - | 1 | MaxPool | - | - |
| CONV-D4 | 4 | - | Conv64D4 | 1 | - | Conv64D4 | - | - | 1 | Conv64D4 |
| CONV-3 | 1 | Conv64 | - | 1 | Conv64 | - | 1 | Conv64 | - | - |
| POOL-3 | 1 | MaxPool | - | 1 | MaxPool | - | 1 | MaxPool | - | - |
| CONV-D8 | 1 | - | Conv128D8 | 1 | - | Conv128D8 | - | - | 1 | Conv128D8 |
| CONV-D4-2 | 4 | - | Conv64D4 | 1 | - | Conv64D4 | - | - | 1 | Conv64D4 |
| CONV-D2-2 | 4 | - | Conv32D2 | 1 | - | Conv32D2 | - | - | 1 | Conv32D2 |
| CONV-4 | 1 | Conv128 | - | 1 | - | Conv16 | - | - | 1 | Conv16 |
| CONV-5 | - | - | - | 1 | Conv128 | - | 1 | Conv128 | - | - |
| CONCATE & CROP | 1 | CONV-4, CONV-D2-2 | | - | - | | - | - | - | - |
| UP-CONV-1 | 1 | Transpose-Conv | | 1 | Transpose-Conv | | 1 | Transpose-Conv | - | - |
| CONCATE & CROP | 1 | UP-CONV-1, CONV-3, CONV-D4-2 | | 1 | UP-CONV-1, CONV-3 | | 1 | UP-CONV-1, CONV-3 | - | - |
| CONV-6 | 1 | Conv64 | | 1 | Conv64 | | 1 | Conv64 | - | - |
| UP-CONV-2 | 1 | Transpose-Conv | | 1 | Transpose-Conv | | 1 | Transpose-Conv | - | - |
| CONCATE & CROP | 1 | UP-CONV-2, CONV-2, CONV-D8 | | 1 | UP-CONV-2, CONV-2 | | 1 | UP-CONV-2, CONV-2 | - | - |
| CONV-7 | 1 | Conv32 | | 1 | Conv32 | | 1 | Conv32 | - | - |
| UP-CONV-3 | 1 | Transpose-Conv | | 1 | Transpose-Conv | | 1 | Transpose-Conv | - | - |
| CONCATE & CROP | 1 | UP-CONV-3, CONV-1, CONV-D2 | | 1 | UP-CONV-3, CONV-1 | | 1 | UP-CONV-3, CONV-1 | - | - |
| CONV-8 | 1 | Conv16 | | 1 | Conv16 | | 1 | Conv16 | - | - |
| CONCATE & CROP | - | - | | 1 | CONV-8, CONV-4 | | - | - | - | - |
| OUTPUT 82 × 82 | 1 | Fully Convolutional Layer | | 1 | Fully Convolutional Layer | | 1 | Fully Convolutional Layer | 1 | Fully Convolutional Layer |

Some random feature maps of each layer in the BowNet network during training process are shown in Figure 5. The sparsity of dilated convolutional layer causes a bigger receptive field whereas the result might be the false detected area as the tongue contour but with more accurate shapes. Checkerboard artifact can be seen clearly in up-sampling layers where deconvolutional layers are applied on the previous feature maps. Summation of both dilated and regular convolution layers results in a uniform contour region in the output results.

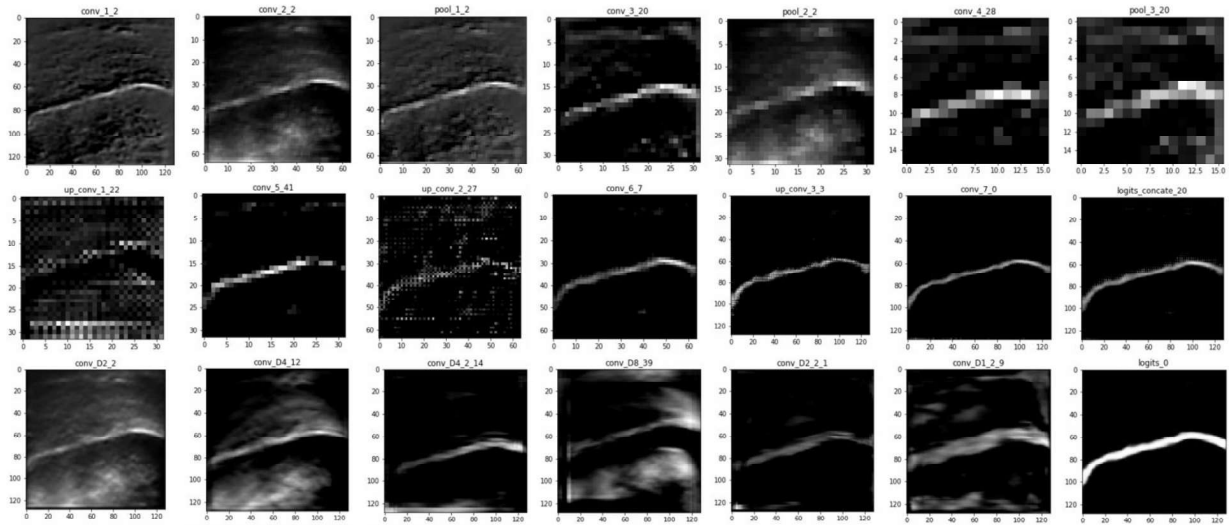

Figure 5. Random feature maps from different layers of the BowNet. Two rows are related to the encoder-decoder forward path, and the last row is related to the dilated consecutive dilated convolutions. Checkerboard effect in the up-sampling section can be seen clearly.

## 4 Experimental Results

### 4.1 System setup

In order to train our proposed network models, first we conducted an extensive random search hyperparameter tuning (Bergstra, James and Bengio, 2012) for finding the optimum value of parameters in each network such as filter size (double for each consecutive layer starts from 16, 32, or 64), kernel size (3 × 3 and 5 × 5), dilation factor (double for each consecutive layer starts from 1 or 2), the number of global iterations (iteration and epoch size), batch size (10, 20, and 50 depend on GPU memory), augmentation parameters (online and offline), drop-out factor (0.5, 0.7, and variable), padding type (zero or valid), normalization layer (with and without), optimization methods (SGD and Adam (Kingma and Ba, 2014)), and the type of the activation layers (tanh and ReLU). We also tested several network configurations for the BowNet whereas the number of layers was different in encoding-decoding forward paths (3, 4, and 5), and for the wBowNet, where the number of dilated convolution layers was different. In this experiment, we set real-time performance as one restriction for all the architectures in this study.

Our results from hyperparameter tuning revealed that, besides network architecture size, learning rate has the most significant effect on the performance of each architecture in terms of accuracy. Testing fixed and scheduled decaying learning rate showed that the variable learning rate might provide better results, but it requires different initialization of decay factor and decay steps. Therefore, for the sake of a fair comparison, we only reported results using fixed learning rates. Table 2. presents selected learning rates for each proposed network model.

Table 2. Fixed Learning-rate (LR) tuning using Best training and validation loss (BTL, BVL).

| LR | sUNet | | sDeepLab | | BowNet | | wBowNet | |
|---|---|---|---|---|---|---|---|---|
| | BTL | BVL | BTL | BVL | BTL | BVL | BTL | BVL |
| 0.005 | N/A | N/A | N/A | N/A | N/A | N/A | N/A | N/A |
| 0.0001 | 0.023 | 0.024 | 0.02 | 0.023 | 0.023 | 0.023 | 0.022 | 0.021 |
| 0.0003 | 0.021 | 0.023 | 0.018 | 0.021 | 0.021 | 0.022 | 0.021 | 0.023 |
| 0.0005 | **0.020** | 0.020 | **0.014** | 0.020 | **0.020** | 0.021 | **0.021** | 0.021 |
| 0.0007 | 0.019 | 0.022 | 0.017 | 0.020 | 0.020 | 0.021 | 0.020 | 0.022 |
| 0.0009 | 0.020 | 0.023 | 0.017 | 0.020 | 0.019 | 0.021 | 0.022 | 0.023 |
| 0.00001 | 0.026 | 0.028 | 0.026 | 0.029 | 0.027 | 0.027 | 0.026 | 0.031 |

Architectures were deployed using the publicly available TensorFlow framework on Keras API as the backend library (Abadi et al., 2016; Chollet and others, 2015). For initialization of networks parameters, randomly distributed values have been selected similarly using the same seed value. For all the four deep convolutional networks, size of the input images, the number of iterations, mini-batch sizes, number of epochs were selected as 128 × 128, 3000, 10, 60, respectively. Adam optimization was selected with a fixed momentum value of 0.9 for finding the optimum solution on a binary cross entropy loss function. Each network model was trained for ten times, then average and the standard deviation were reported. It takes less than one hour approximately for each time training of a model depend on the network size, using one NVIDIA 1080 GPU unit which was installed on a Windows PC with Core i7, 4.2 GHz speed, and 32 GB of RAM. We also used the Google cloud virtual machine with a Tesla P100 GPU and 16GB of memory for acquiring training results faster.

## 4.2 Datasets and Data Augmentation

We evaluated our proposed deep network models using different scenarios (train and test on one dataset, train on one and test on another dataset, and train and test on both datasets). The two datasets comprise of ultrasound tongue images acquired with different ultrasound machines: one from University of Ottawa (Dataset I) and another from the publicly available SeeingSpeech project (Dataset II) (Lawson et al., 2015). True labels corresponding to each image were created by two experts manually using our customized annotation software. Users should only annotate several point markers on the edge of the tongue contour region, and then the contour curve will be created using the B-spline method automatically between point markers.

As Table 3 shows, we separated datasets into training, validation, and test sets. Each model was trained and validated on the dataset from online augmentation, and then it was tested on both datasets separately. Because we used the same batch size, the number of iterations, the number of epochs, and the equal initial randomization, each model will see the same set of images during the training, validation, and testing procedure. Note that there is a corresponding annotated mask (ground truth) for each image in both datasets which is augmented online, correspondingly. As Figure 9. shows, ranking each image in terms of the distance from the average image, dataset II has similar images (homogenous) in comparison to the dataset I (heterogeneous). Figure 6. shows two samples of Database I and II with their corresponding ground truth labels.

Table 3. Specification of the dataset I and II for training using online augmentation

|  | Total number of images | Train/Validation (%80/10) | Test (%10) |
|---|---|---|---|
| Dataset I (UOttawa) | 2058 | 1646/205 | 205 |
| Dataset II (SeeingSpeech) (Lawson et al., 2015) | 4016 | 3212/401 | 401 |

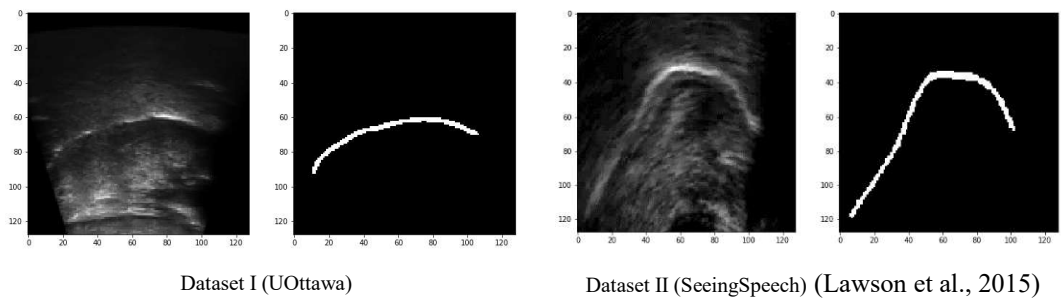

Dataset I (UOttawa)    Dataset II (SeeingSpeech) (Lawson et al., 2015)

Figure 6. Two sample images from Dataset I and II accompany with their corresponding truth labels.

Online augmentation on ultrasound data was deployed using the Keras augmentation library with the same random numbers for the training of each deep convolutional network. In total, 60 iterations × 10 mini-batches × 50 epochs = 30000 augmented images were used for training and validation of each network models. We used image flipping (half of the dataset after augmentation), rotation (50-degree rotation in each side) and zooming (ratio of 0.5x to 1.5x) to mimic all the possible transformation which can happen in ultrasound tongue data. Some randomly selected

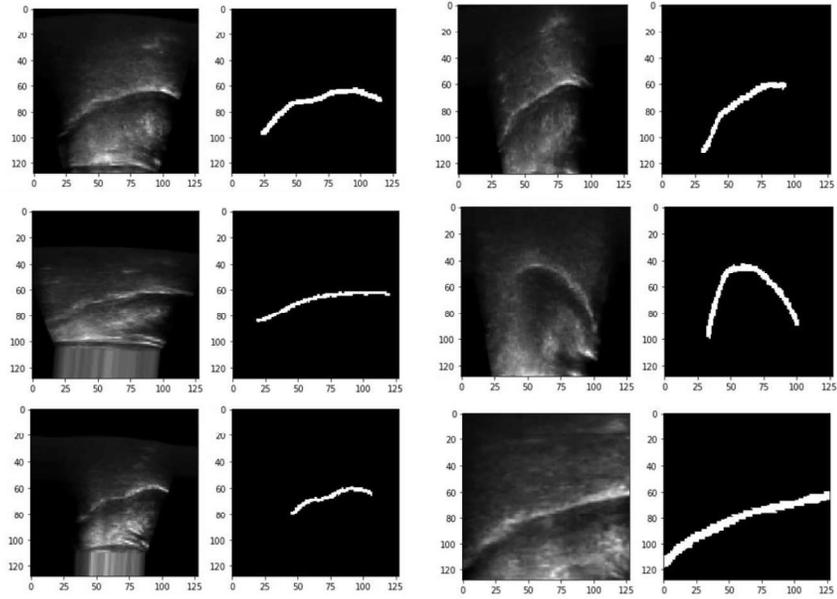

augmented data are presented in

Figure 7. As we can see clearly in the figure, after applying augmentation on truth labels, due to the interpolation for down- or up-sampling in the zooming process, they become gagged after binarization. Note that, the correct truth labels for this application from semantic segmentation literature should be a binary image instead of gray-scale. Furthermore, our experiments show that the multiplication of the rotation matrix will slightly change intensity values due to the approximation in calculations. Moreover, some artifacts might be added to the data due to the online

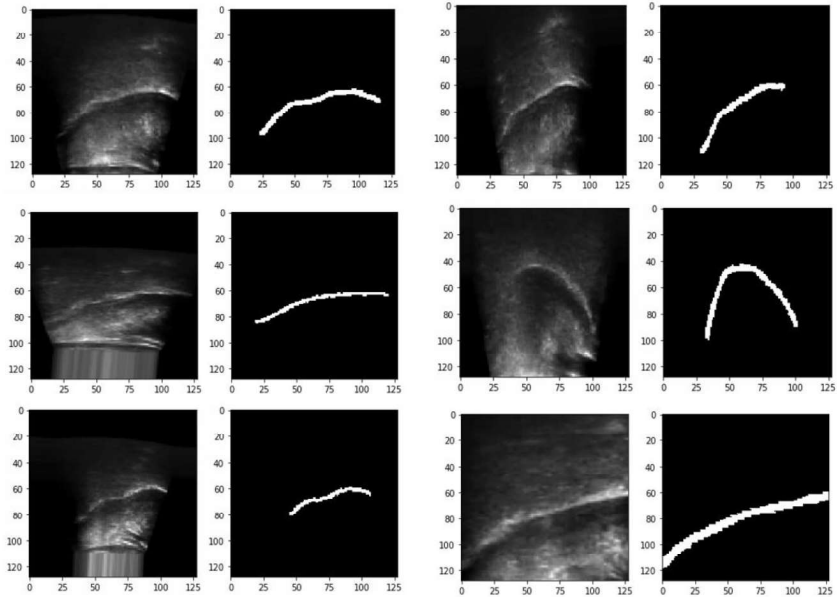

augmentation (see

Figure 7 lower left column).

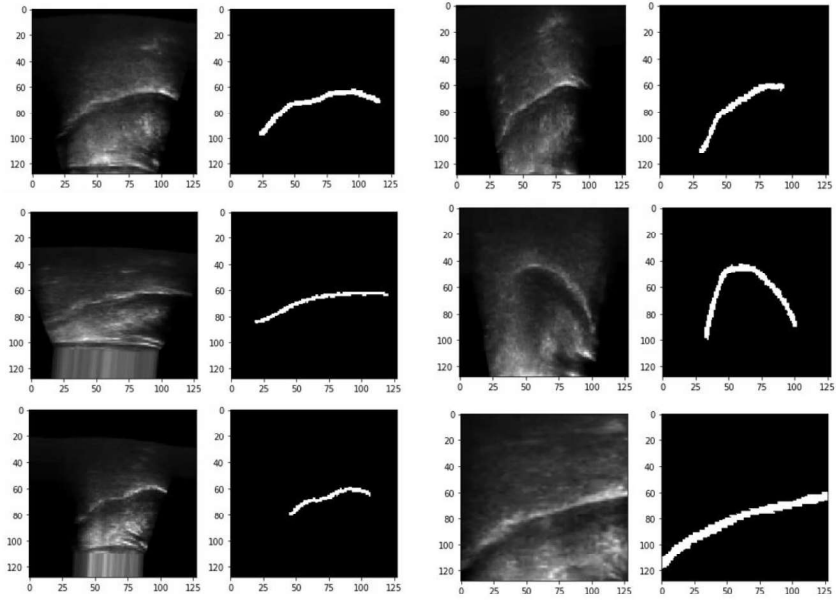

Figure 7. Some randomly selected images after online augmentation during the training process.

For this reason, we created an enhanced dataset using offline augmentation to see the effect of those differences in the dataset on the performance of each deep learning model. We re-examined all experiments on two enhanced datasets from the dataset I and II. Due to the heterogeneity and homogeneity of the dataset I and II, we applied the informed undersampling method for imbalanced datasets (Liu et al., 2006) to create two similar datasets with the highest variations for better regularization and better comparison. For this reason, we first find the average image of each dataset, and then the Euclidian distance of images was calculated with respect to the average image. Figure 8. shows the average image of each dataset as well as a sample image with its difference from the average one.

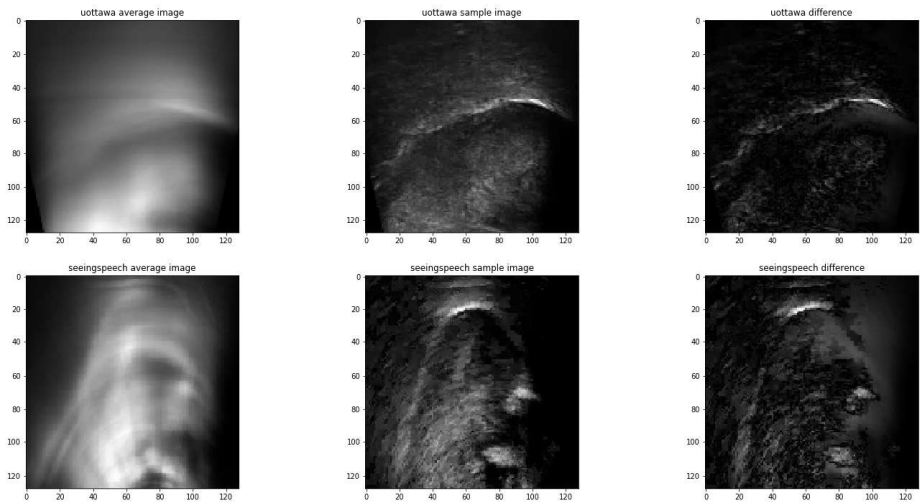

Figure 8. Informed under-sampling procedure to enhance dataset in terms of variation: Left column: the average image from Dataset I and II. Middle column: one sample image from each dataset. The right column: the difference between the sample image and the average one.

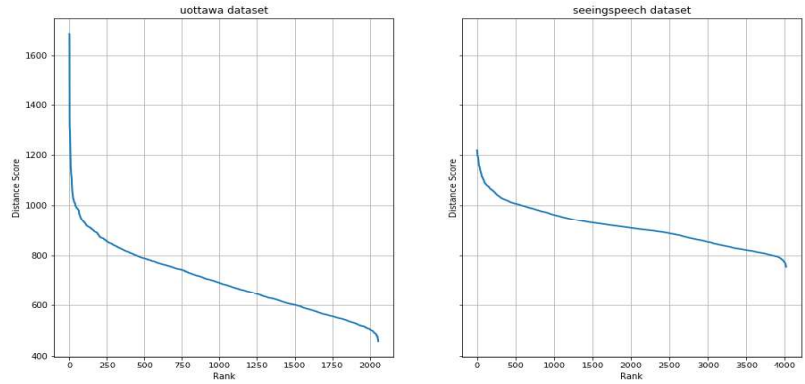

Figure 9. The ranking dataset I and II in terms of similarity distance score calculated the distance from each image to the average image.

After sorting each dataset in terms of distance score, we selected 2000 images with the highest rank and 50 images from the lowest position. Then, we applied offline augmentation using a publicly available augmentation software Augmenter (Bloice et al., 2017). We applied flipping, rotation, and zooming (with the same parameters as online augmentation) to mimic all the possible transformation which can happen in ultrasound data of the tongue. As can be seen from Figure 10 and Figure 11, two datasets are similar and heterogeneous after offline augmentation. To alleviate the problems of gagged images using online augmentation, we applied gaussian blurring on ground truth labels followed by binarized and morphological erosion and closing operations.

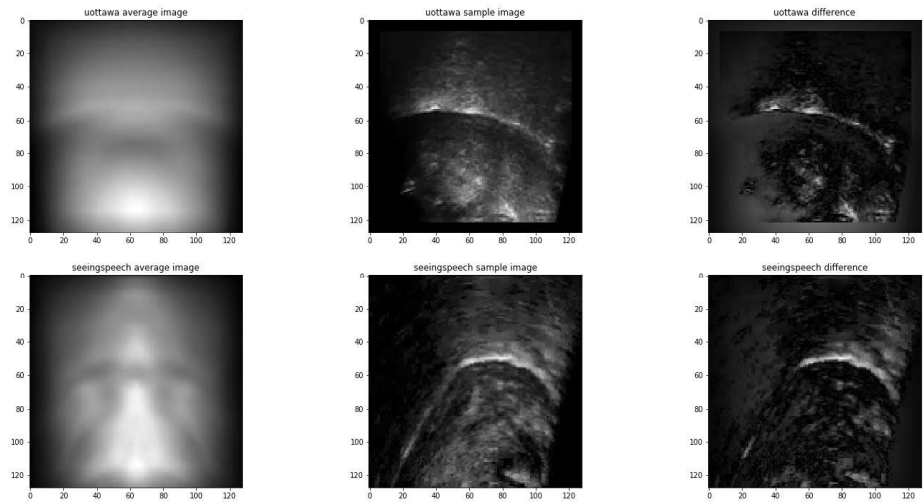

Figure 10. After informed undersampling and offline augmentation: the average image from Dataset I and II. Middle column is one sample image from each dataset. The right column shows the difference between the sample image from the average one.

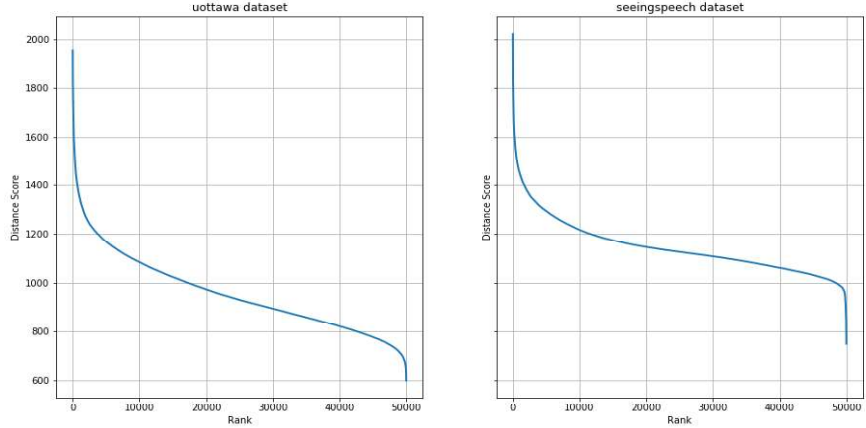
Figure 11. The ranking dataset I and II in terms of similarity distance score calculated the distance from each image to the average image after applying informed undersampling and offline augmentation.

Table 4 shows the specification of datasets I and II after offline augmentation and enhancement. We also created a new dataset using a combination of the two datasets with 100000 images and with the same train/test ratio.

Table 4. Datasets information after informed undersampling and offline augmentation.

|  | Total number of images | After informed undersampling | After offline augmentation and enhancement | Training/Validation (%90/%5) | Testing (5%) |
|---|---|---|---|---|---|
| Dataset I (UOttawa) | 2058 | 2050 | 50000 | 45000/2500 | 2500 |
| Dataset II (SeeingSpeech) | 4016 | 2050 | 50000 | 45000/2500 | 2500 |

## 4.3 Experiments and Validation

### 4.3.1 Qualitative Analysis

To illustrate the efficacy of our proposed segmentation methods qualitatively, some randomly selected segmentation results from testing our proposed models on datasets I and II using online and offline augmentation are shown in this section. Figure 12. shows the results from applying of each proposed model on both datasets using online augmentation. As it can be seen clearly, wBowNet could achieve better prediction results in terms of noise. Training and testing on the same dataset using online augmentation, all the models in this study generate instances with acceptable noise. On the case of training on one dataset and testing on another one results show a worse prediction for both sDeepLab and sUNet. It is noteworthy to mention that the correct position of the tongue contour curve acquired from skeletonizing should be shifted toward the edge of the intensity gradient between black and white regions (Stone, 2005). In these experiments, we ignored that moving to make a comparison study the same for all the models.

Segmentation results of the proposed models on datasets after enhancement and using offline augmentation is presented in Figure 13. Similar to the previous experiment, each model was trained on one dataset and tested on both datasets separately. To extract tongue contours, form the segmented region, we had to enhance the prediction map before applying the skeletonizing method for having a fair comparison. For this reason, first, the biggest object in prediction image was selected using areas calculation of objects (see the second rows of Figure 13) and then skeletonization method was applied on the enhanced predicted map. The generated contours are compared with the skeletonized contour of the ground truth labels.

From Figure 13, we can see that the wBowNet could predict better feature maps with less noise and false prediction regions than other models. The last row of figures shows the difference between the prediction curve and the truth label curve. Faded area means a better correlation between the contour from the ground truth label and predicted map. Unlike the Figure 12, it can be seen from the first and second rows of Figure 13 that the prediction maps are almost binary images instead of grayscale since the ground truth labels are binary with two classes of tongue contour and background regions following the semantic segmentation literature.

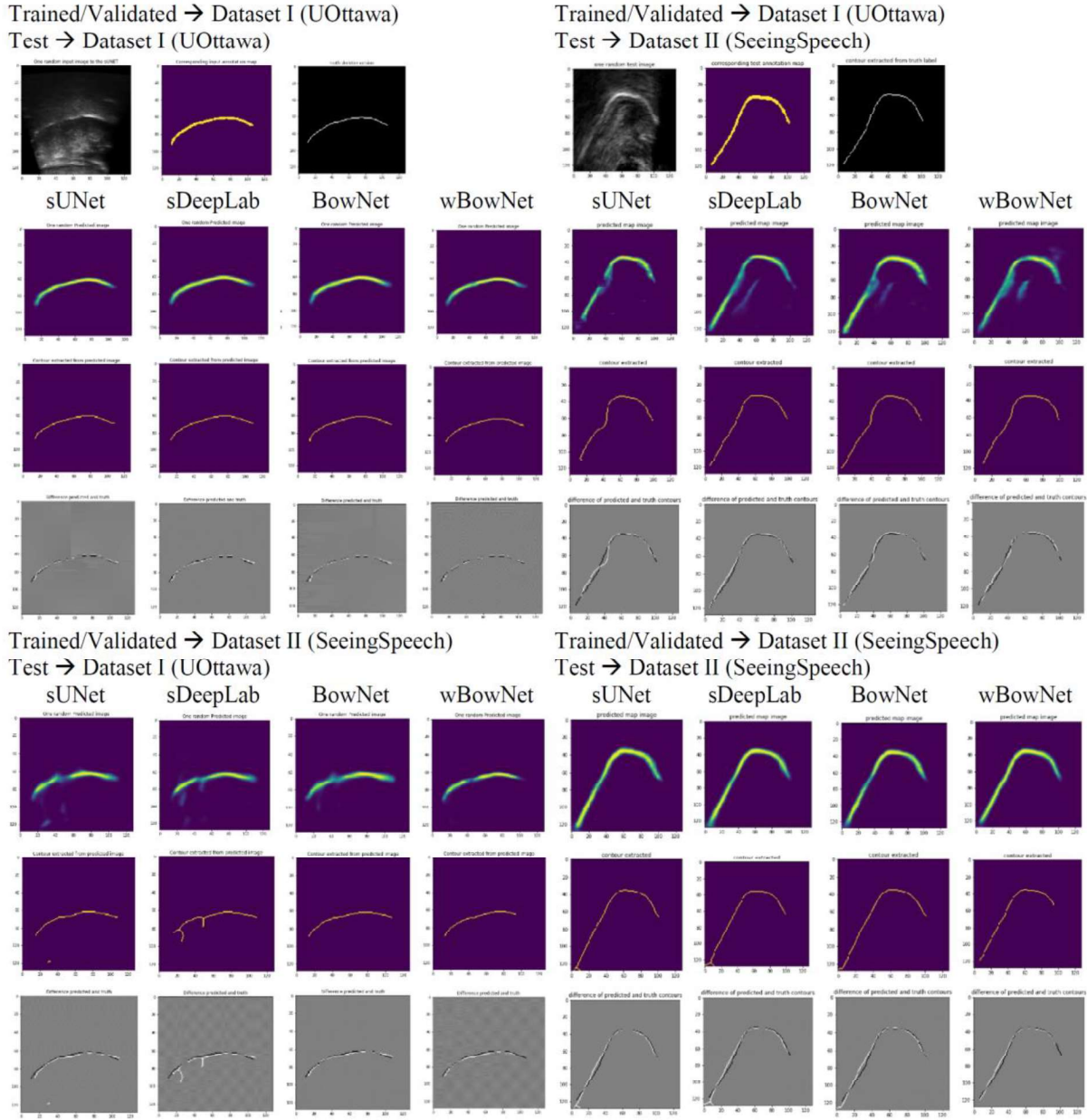

Figure 12. Sample test results of trained and validated models on datasets using online augmentation. First rows are prediction maps where yellow regions are a higher probability of true prediction than other areas. Second rows are contour extracted from predicted feature maps after the same binarization and skeletonization. Last rows are the difference between the contour extracted from the true label and predicted label. Note that the sample image, ground truth, and its contour are illustrated on the top of the first row.

As it can be seen clearly in the last column of Figure 13, both BowNet models could obtain outstanding prediction maps in comparison with other architectures. From the same figure, although BowNet models had significantly better outcomes in both experiments, the difference between contours was comparable for all the network models. For instance, sDeepLab has acceptable difference images while prediction maps contain more noise than other models (see the last rows of Figure 13). Therefore, due to the post-processing stages on prediction map, sUNet and sDeepLab obtained comparable results like BowNet models in the quantitative study in the next section although they could not generate similar prediction maps. Therefore, the quantitative research in this work is a support for the qualitative study, and the results of both evaluations should be considered together.

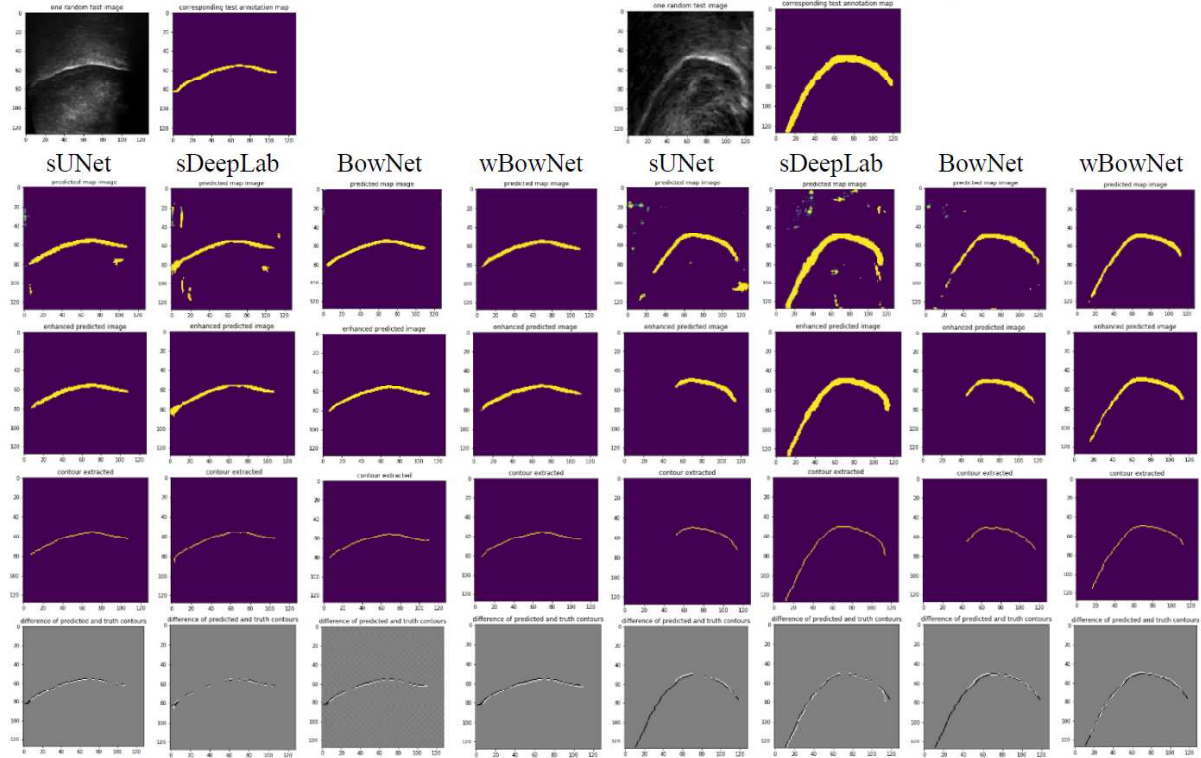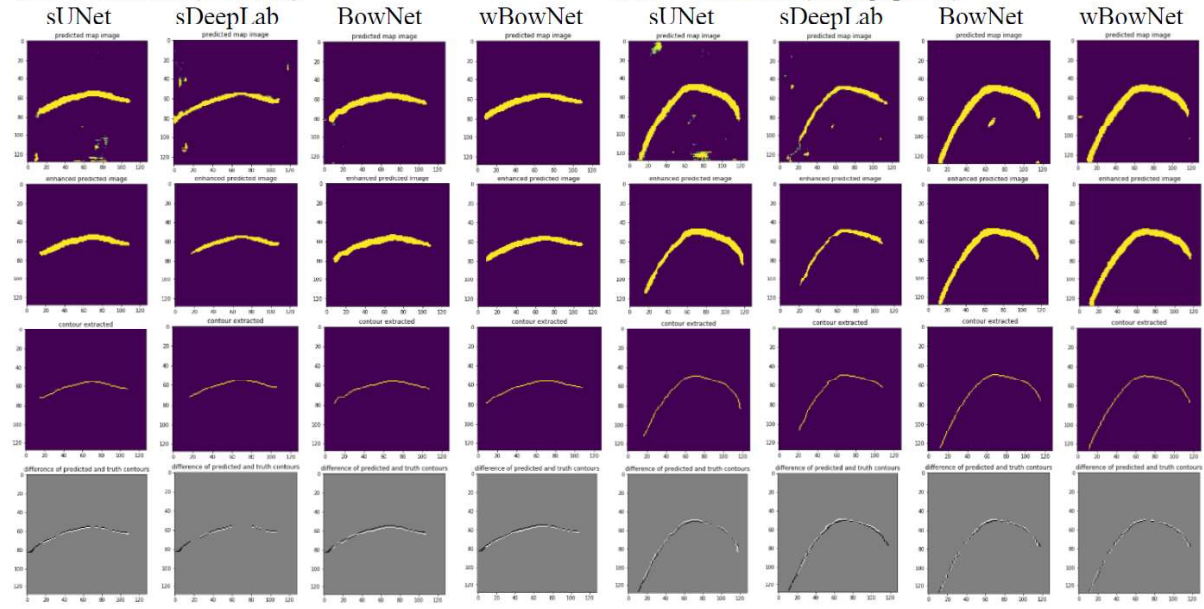

Figure 13. Sample test results of trained and validated models on enhanced datasets using offline augmentation. First rows are prediction maps where yellow regions are a higher probability of true prediction than other areas. Second rows are enhanced prediction maps using truncated smaller areas. Third rows are contour extracted from predicted feature maps after binarization and skeletonization. Last rows are the difference between the contour extracted from ground true labels and predicted label. Note that the sample image and its corresponding ground truth of each dataset are illustrated on the top of the first row.

We also trained and validated each model on a dataset which comprises of data from both dataset I and II using offline augmentation and enhancement. Then, each model tested on both datasets separately like the previous experiments. Figure 14 shows the consistency of our last conclusion about BowNet and wBowNet on the bigger dataset. As it can be seen clearly from the figure, both BowNet architectures could predict segmented images with less noise than other models. Difference images of tongue contours for sUNet and sDeepLab are analogous to the BowNet networks after enhancement of the feature maps while those models could not achieve better predictions than BowNet and wBowNet.

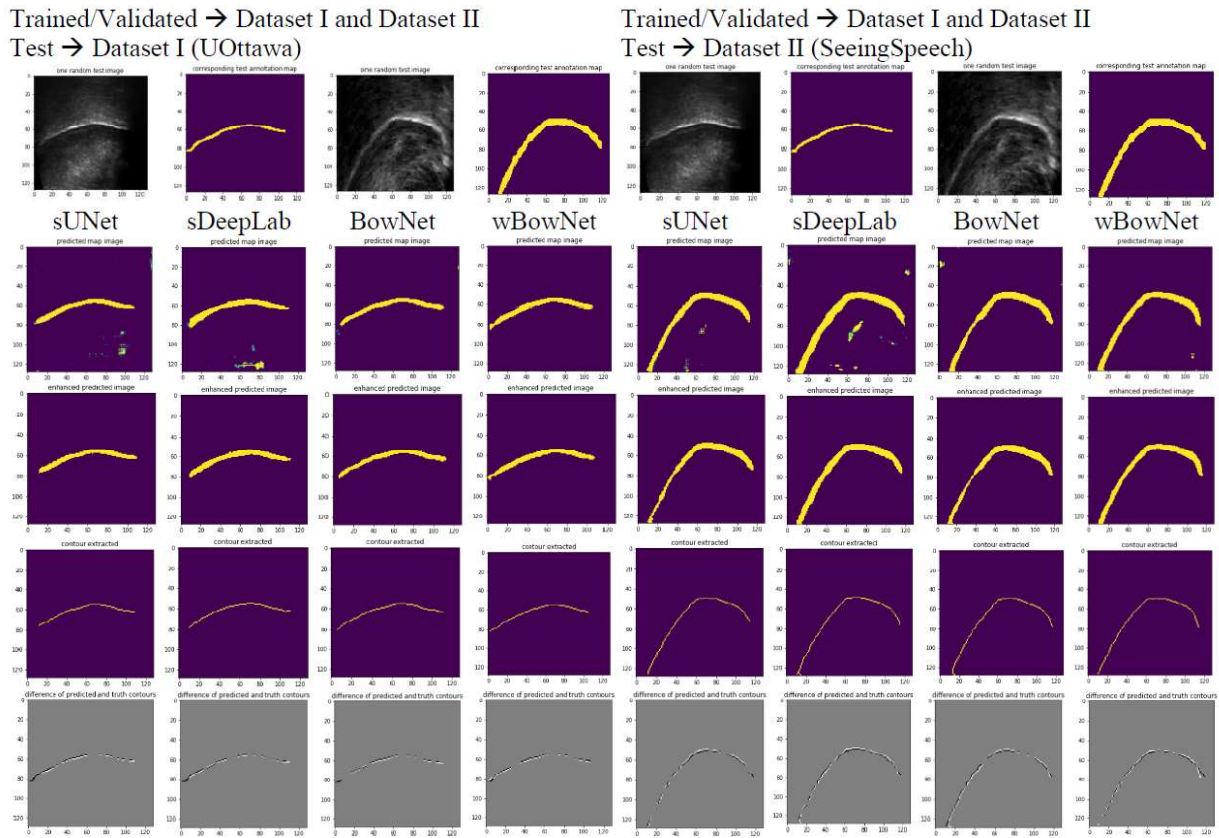

Figure 14. Sample test results of trained and validated models on the combination of two enhanced datasets using offline augmentation. The first row is prediction maps where yellow regions are a higher probability of true prediction than other areas. The second row is enhanced prediction maps. The third row is contour extracted from the feature predicted feature map after binarization and skeletonization. Last rows are the difference between the contour extracted from the true label and predicted label. Note that the sample image and ground truth of each dataset are illustrated on top of the first row.

### 4.3.2 Quantitative Evaluation and Comparison Study

In order to quantitatively validate our proposed models, we followed the standard evaluation criteria for the field of ultrasound tongue segmentation (Akgul et al., 1999; Fasel and Berry, 2010; Hahn-Powell et al., 2014; Laporte and Ménard, 2018; Milletari et al., 2016; Xu et al., 2016a). Contours extracted from prediction maps and ground truth labels are compared in terms of the mean sum of distance (MSD) as defined in equation (1) (Jaumard-Hakoun et al., 2016). It provides an evaluation metric defined as the mean distance between pixels of a contour $U$ and a contour $V$, even if these curves do not share the same coordinates on one axis or do not have the same number of points. The latter one consisted of a contour set $U$ of 2D points $(u_1, \cdots, u_n)$ and the former one consisted of a contour set $V$ of 2D points $(v_1, \cdots, v_m)$.

We carried out contour extraction using the skeleton method on binarized ground truth labels and predicted images employing morphological operators from the Scikit-learn library (Martínková et al., 2004). This criterion is sensitive to translation as we saw in our experiments. For this reason, we did not shift the extracted contours to their correct positions for the sake of a fair comparison.

$$MSD(U,V) = \frac{1}{m+n}\left(\sum_{i=1}^{m} \min_{j} |v_i - u_j| + \sum_{i=1}^{n} \min_{j} |u_i - v_j|\right) \quad (1)$$

To evaluate the performance of the proposed methods, we also calculated and reported the value of the dice coefficient and binary cross-entropy loss between non-enhanced prediction maps and labeled data. Binary cross-entropy loss which is a particular case of multiclass cross-entropy calculates the difference between the predicted and the ground truth labeled data, as defined in equation (2). For assessment of the proposed method performance, we also calculated the dice coefficient as described in equation (3) where $p_i \in [0, 1]$ is the $i^{th}$ predicted pixel in the flattened image, whereas the original pixel value from [0, 255] in a grayscale image is rescaled to [0, 1] for fitting into the ReLU/sigmoid activation functions. And $y_i \in [0, 1]$ is the corresponding class label for that $i^{th}$ pixel of the sample data.

$$L_{BCE} = \sum_i y_i \log p_i + (1 - y_i)\log(1 - p_i) \quad (2)$$

$$L_{Dice} = -\frac{2\sum_i p_i y_i}{\sum_i p_i + \sum_i y_i} \quad (3)$$

Quantitative results are reported from the same experiments in the qualitative study with the same data, setups, and procedures. Table 5. shows the results of training and validating each deep network models on dataset I and II using online augmentation. Each architecture was tested on both datasets separately. For each experiment, we run each system ten times, and then average and standard deviation over the results are tabulated. Last results are only presented to assess the robustness of each model in terms of overfitting and convergence. Therefore, analogy between the last training and validation losses means better convergence. The best results in tables ascertain the accuracy of each model in terms of loss function and dice coefficient. The results of Table 5 reveals that the BowNet and wBowNet could reach optimum solutions better than other models. From Figure 15, we can see that all the deep learning models trained and validated datasets with the same trend while wBowNet has the lowest fluctuations. One reason for significant volatility in training and validation trend is the effect of drop-out whereas the model cannot generalize some mini batches when random parameters are deactivated continuously. BowNet model has the second place in terms of robustness as it can be seen in Table 5.

Table 5. also shows the number of parameters and the memory which is needed to save those trainable parameters, calculated by TensorFlow library. As can be seen from the table, BowNet has the lowest number of parameters. wBowNet and sDeepLab have almost double the number of parameters of BowNet. As an example, we calculated the number of parameters of the original UNet as 31042369 with the similar method for our proposed architectures which is 71 times bigger than BowNet and 39 times bigger than wBowNet.

Table 5. Results of the best and the last binary cross entropy loss function and dice coefficient.

| Trained/Validated → Dataset I (UOttawa) | | sUNet | | sDeepLab | | BowNet | | wBowNet | |
|---|---|---|---|---|---|---|---|---|---|
| Number of parameters/ Memory intake (KB) | | 948833 | 3795332 | 785889 | 3143556 | **434785** | **1739140** | 786657 | 3146628 |
| | | Mean | STD | Mean | STD | Mean | STD | Mean | STD |
| Train loss | Last | 0.0242 | 0.2329 | 0.0336 | 0.0393 | 0.0265 | 0.0427 | **0.0242** | **0.037** |
| Validation loss | Last | 0.0321 | 0.0075 | 0.0367 | 0.0131 | **0.0254** | **0.0046** | 0.0306 | 0.0103 |
| Train Dice | Last | 0.7367 | 0.2329 | 0.8064 | 0.0393 | 0.7871 | 0.0427 | **0.817** | **0.037** |
| Validation Dice | Last | 0.8061 | 0.0277 | 0.796 | 0.0222 | 0.7886 | 0.0389 | **0.8068** | **0.016** |
| Train loss | Best | 0.0109 | 0.0003 | 0.0111 | 0.0003 | 0.0107 | 0.0005 | **0.0104** | **0.0005** |
| Validation loss | Best | 0.0123 | 0.0012 | 0.0115 | 0.001 | **0.0109** | **0.0007** | 0.0122 | 0.0019 |
| Train Dice | Best | 0.8751 | 0.0055 | 0.8775 | 0.002 | 0.8748 | 0.0053 | **0.8846** | **0.0142** |
| Validation Dice | Best | 0.8582 | 0.0036 | 0.8594 | 0.006 | 0.859 | 0.0058 | **0.8652** | **0.0051** |
| Trained/Validated → Dataset II (SeeingSpeech) | | sUNet | | sDeepLab | | BowNet | | wBowNet | |
| | | Mean | STD | Mean | STD | Mean | STD | Mean | STD |
| Train loss | Last | 0.0512 | 0.053 | 0.0458 | 0.0669 | 0.0592 | 0.0269 | **0.045** | **0.0454** |
| Validation loss | Last | 0.037 | 0.007 | 0.0432 | 0.0113 | 0.0464 | 0.0118 | **0.0448** | **0.0162** |
| Train Dice | Last | 0.7688 | 0.053 | 0.7774 | 0.0669 | 0.7746 | 0.0269 | **0.8024** | **0.0454** |
| Validation Dice | Last | 0.7932 | 0.0316 | 0.7918 | 0.0239 | 0.7864 | 0.0316 | **0.7912** | **0.0312** |
| Train loss | Best | 0.0242 | 0.0004 | 0.0224 | 0.0009 | 0.0242 | 0.0008 | **0.0198** | **0.0004** |
| Validation loss | Best | 0.0258 | 0.0008 | 0.0234 | 0.0019 | 0.0258 | 0.0008 | **0.0228** | **0.0013** |
| Train Dice | Best | 0.863 | 0.0053 | 0.8608 | 0.0004 | 0.8658 | 0.0026 | **0.8698** | **0.0008** |
| Validation Dice | Best | 0.852 | 0.0041 | 0.86 | 0.0043 | 0.85 | 0.0099 | **0.865** | **0.0058** |

The results of testing each model using MSD criteria applying on the contours extracted from enhanced prediction maps and the ground truth labels are illustrated in Table 6. As we discussed before due to removing noise from model's instances during enhancement procedure, wBowNet is not the best model in terms of accuracy although it outperformed other network models regarding MSD criteria and qualitative results in many cases. To convert MSD from pixel to millimeter, we calculated the approximate conversion of 1 px = 0.638 mm for both datasets I and II. wBowNet and BowNet could achieve MSDs around 0.04mm on average while the Deep Belief Network (DBN) model in a study by (Fasel and Berry, 2010) achieved 1.0 mm (1 px = 0.295 mm). For the active contour models mentioned in (Li et al., 2005), the average MSD was 1.05 mm. It is important to mention that the two human experts participating in active contour models experiment produced two different annotation results having an average MSD of 0.73 mm (Li et al., 2005), which might thus be reasonably considered the ultimate approximate minimum MSD value based human capability.

Table 6. Results of testing each trained model using online augmentation on different datasets created. MSD is calculated and converted into millimeters with approximation depend on the ultrasound device resolution.

| Trained/Validated → Dataset I (UOttawa) | | | | |
|---|---|---|---|---|
| Test → Dataset I (UOttawa) | sUNet | sDeepLab | BowNet | wBowNet |
| Test loss | 0.02361 | 0.02711 | 0.02888 | **0.01896** |
| Test Dice | 0.75985 | 0.74899 | 0.74466 | **0.82629** |
| MSD (pixels) | 0.2496 | 0.2666 | 0.2819 | **0.2136** |
| MSD (mm) | 0.03744 | 0.0399 | 0.0422 | **0.0320** |
| Test → Dataset II (SeeingSpeech) | | | | |
| Test loss | 0.0829 | **0.06475** | 0.06964 | 0.06574 |
| Test Dice | 0.4985 | 0.52630 | **0.54718** | 0.52004 |
| MSD (pixels) | 0.4249 | **0.2408** | 0.3742 | 0.3588 |
| MSD (mm) | 0.0637 | **0.0361** | 0.0561 | 0.0538 |
| Trained/Validated → Dataset II (SeeingSpeech) | | | | |
| Test → Dataset II (SeeingSpeech) | | | | |
| Test loss | 0.04876 | 0.04540 | 0.04522 | **0.04367** |
| Test Dice | 0.71770 | 0.73029 | 0.71426 | **0.73680** |
| MSD (pixels) | 0.38279 | 0.39123 | 0.40143 | **0.26976** |
| MSD (mm) | 0.05741 | 0.05868 | 0.06021 | **0.04046** |
| Test → Dataset I (UOttawa) | | | | |
| Test loss | 0.04400 | 0.04222 | **0.03948** | 0.06419 |
| Test Dice | 0.57989 | 0.57879 | **0.62013** | 0.39702 |
| MSD (pixels) | 0.26377 | 0.37512 | 0.2874 | **0.18037** |
| MSD (mm) | 0.03956 | 0.05626 | 0.04311 | **0.02705** |

We applied deep network models again on enhanced datasets which were created using offline augmentation. As Table 7 shows, the sDeepLab has the best training/validating accuracy, but from the qualitative study, we saw that instances of the sDeepLab contain more false prediction regions than others. The reason for this contradiction can be seen in

Table 8 where the sDeepLab has the worse instance prediction values due to the overfitting. From the Table 7 and

Table 8, wBowNet achieved distinguished results in terms of accuracy for test datasets, as we could anticipate that from the qualitative study, whereas results of training/validation show a better convergence and overfitting.

From

Table 8, although each architecture has its own best performance but the wBowNet could obtain results with smallest difference with the best architecture in each experiment. For instance, in two experiments, wBowNet achieved the best MSD values and in other two experiments, its MSD results has only about 0.002mm difference with the best architecture. Alternatively, sUNet could reach to the best MSD values in two experiments but in two others its difference with the best model was 0.01mm. Note that the number of parameters for sUNet in this example is around %20 more than wBowNet.

Table 7. Results of the best and the last binary cross entropy loss function and dice coefficient for enhanced datasets from using offline augmentation. The number of parameters along with their memory intake is also illustrated in the second row.

| Trained/Validated → Dataset I (UOttawa) | | sUNet | | sDeepLab | | BowNet | | wBowNet | |
|---|---|---|---|---|---|---|---|---|---|
| Number of parameters / Memory intake (KB) | | 948833 | 3795332 | 785889 | 3143556 | **435233** | **1740932** | 794849 | 3179396 |
| | | Mean | STD | Mean | STD | Mean | STD | Mean | STD |
| Training Loss | Last | 0.0322 | 0.0064 | 0.0318 | 0.0073 | 0.0346 | 0.0072 | **0.0348** | **0.0053** |
| Validation Loss | Last | 0.0375 | 0.0041 | 0.0336 | 0.0061 | 0.0381 | 0.0064 | **0.0361** | **0.0057** |

| | | | | | | | | |
|---|---|---|---|---|---|---|---|---|
| Training Dice | Last | 0.7840 | 0.0210 | 0.8234 | 0.0196 | 0.7785 | 0.0340 | **0.7660** | **0.0332** |
| Validation Dice | Last | 0.7593 | 0.0263 | 0.7983 | 0.0250 | 0.7693 | 0.0303 | **0.7757** | **0.0247** |
| Training Loss | Best | 0.0196 | 0.0007 | **0.0170** | **0.0019** | 0.0196 | 0.0008 | 0.0228 | 0.0023 |
| Validation Loss | Best | 0.0221 | 0.0009 | **0.0199** | **0.0022** | 0.0221 | 0.0009 | 0.0241 | 0.0021 |
| Training Dice | Best | 0.8524 | 0.0030 | **0.8729** | **0.0137** | 0.8510 | 0.0048 | 0.8360 | 0.0132 |
| Validation Dice | Best | 0.8381 | 0.0066 | **0.8590** | **0.0102** | 0.8406 | 0.0072 | 0.8247 | 0.0124 |

| Trained/Validated → Dataset II (SeeingSpeech) | | sUNet | | sDeepLab | | BowNet | | wBowNet | |
|---|---|---|---|---|---|---|---|---|---|
| | | Mean | STD | Mean | STD | Mean | STD | Mean | STD |
| Training Loss | Last | 0.0358 | 0.0066 | 0.0358 | 0.0132 | 0.0368 | 0.0053 | **0.0418** | **0.0065** |
| Validation Loss | Last | 0.0454 | 0.0046 | 0.0374 | 0.0059 | 0.0344 | 0.0067 | **0.0414** | **0.0035** |
| Training Dice | Last | 0.8296 | 0.0321 | 0.8372 | 0.0287 | 0.7960 | 0.0599 | **0.8122** | **0.0334** |
| Validation Dice | Last | 0.7914 | 0.0584 | 0.8114 | 0.0427 | 0.8082 | 0.0383 | **0.8232** | **0.0251** |
| Training Loss | Best | 0.0224 | 0.0021 | **0.0190** | **0.0028** | 0.0212 | 0.0028 | 0.0222 | 0.0008 |
| Validation Loss | Best | 0.0248 | 0.0026 | **0.0212** | **0.0030** | 0.0244 | 0.0024 | 0.0254 | 0.0015 |
| Training Dice | Best | 0.9034 | 0.0547 | 0.8930 | 0.0206 | 0.8750 | 0.0316 | **0.9110** | **0.0025** |
| Validation Dice | Best | 0.8796 | 0.0293 | 0.8794 | 0.0252 | 0.8630 | 0.0333 | **0.9014** | **0.0026** |

Table 8. Results of testing each model on different enhanced datasets created from offline augmentation when each trained on the same or different dataset. MSD is calculated and converted into millimeters with approximation depend on the ultrasound device resolution.

| | sUNet | | sDeepLab | | BowNet | | wBowNet | |
|---|---|---|---|---|---|---|---|---|
| Trained/Validated → Dataset I (UOttawa) <br> Test → Dataset I (UOttawa) | Mean | STD | Mean | STD | Mean | STD | Mean | STD |
| Test loss | **0.1447** | 0.0326 | 0.2307 | 0.0533 | 0.3242 | 0.2564 | 0.1546 | 0.0396 |
| Test Dice | **0.7617** | 0.0672 | 0.6284 | 0.0543 | 0.6243 | 0.1690 | 0.7464 | 0.0688 |
| MSD (pixels) | **0.2563** | 0.0217 | 0.2619 | 0.0302 | 0.2877 | 0.0549 | 0.2665 | 0.0168 |
| MSD (mm) | **0.0384** | 0.0033 | 0.0392 | 0.0045 | 0.0431 | 0.0082 | 0.0399 | 0.0025 |
| Trained/Validated → Dataset I (UOttawa) <br> Test → Dataset II (SeeingSpeech) | Mean | STD | Mean | STD | Mean | STD | Mean | STD |
| Test loss | 0.5427 | 0.0873 | 0.5894 | 0.0907 | **0.5378** | 0.1697 | 0.5513 | 0.1325 |
| Test Dice | 0.5815 | 0.0868 | 0.5554 | 0.0732 | **0.6216** | 0.0728 | 0.5733 | 0.1359 |
| MSD (pixels) | 0.4701 | 0.0217 | 0.4639 | 0.0236 | 0.4493 | 0.0516 | **0.4159** | 0.0684 |
| MSD (mm) | 0.0705 | 0.0033 | 0.0695 | 0.0035 | 0.0674 | 0.0078 | **0.0623** | 0.0102 |
| Trained/Validated → Dataset II (SeeingSpeech) <br> Test → Dataset II (SeeingSpeech) | Mean | STD | Mean | STD | Mean | STD | Mean | STD |
| Test loss | 0.3255 | 0.1070 | 0.4787 | 0.1825 | 0.3796 | 0.1652 | **0.1992** | 0.0414 |
| Test Dice | 0.7514 | 0.0907 | 0.6389 | 0.1541 | 0.7313 | 0.1324 | **0.8729** | 0.0244 |
| MSD (pixels) | 0.4145 | 0.0986 | 0.4024 | 0.0663 | 0.4060 | 0.1266 | **0.3300** | 0.1033 |
| MSD (mm) | 0.0621 | 0.0148 | 0.0603 | 0.0099 | 0.0609 | 0.0190 | **0.0494** | 0.0155 |
| Trained/Validated → Dataset II (SeeingSpeech) <br> Test → Dataset I (UOttawa) | Mean | STD | Mean | STD | Mean | STD | Mean | STD |
| Test loss | 0.2524 | 0.1211 | 0.3900 | 0.2290 | **0.2227** | 0.1118 | 0.2787 | 0.0193 |
| Test Dice | 0.6795 | 0.1035 | 0.5213 | 0.1715 | **0.6947** | 0.0831 | 0.6646 | 0.0148 |
| MSD (pixels) | **0.2628** | 0.0233 | 0.2659 | 0.0208 | 0.2669 | 0.0077 | 0.2761 | 0.0169 |
| MSD (mm) | **0.0394** | 0.0035 | 0.0399 | 0.0031 | 0.0400 | 0.0012 | 0.0414 | 0.0025 |

As Table 9 shows, BowNet and wBowNet could find better MSD values in comparison to other models as well as better convergence when training and testing is done on a big dataset comprises of both dataset I and II. Although sUNet and sDeepLab reach to better loss function values during the training and testing stage but again from our qualitative study, their results contain artifacts and false predictions.

Table 9. Results of testing each model on a combination dataset from the dataset I and II where offline augmentation was used for the creation of data. Each trained model was tested in separate experiments on the test data from the dataset I and II. MSD is calculated and converted into millimeters with approximation depend on the ultrasound device resolution. UNet results are reported as the average of 5 times execution.

| | | sUNet | | sDeepLab | | BowNet | | wBowNet | |
|---|---|---|---|---|---|---|---|---|---|
| | | Ave | Sd | Ave | Sd | Ave | Sd | Ave | Sd |
| Training Loss | Last | 0.0416 | 0.001949 | 0.0350 | 0.007176 | 0.0434 | 0.005983 | **0.0452** | 0.010849 |
| Validation Loss | Last | 0.0446 | 0.007503 | 0.0408 | 0.009066 | 0.0472 | 0.009497 | **0.0472** | 0.011606 |
| Training Dice | Last | 0.8082 | 0.043774 | 0.8078 | 0.031076 | 0.7900 | 0.041152 | **0.7768** | 0.028560 |
| Validation Dice | Last | 0.7924 | 0.037819 | 0.7908 | 0.035773 | 0.7540 | 0.056996 | **0.7744** | 0.059383 |
| | | | | | | | | | |
| Training Loss | Best | 0.0230 | 0.000707 | 0.0221 | 0.000707 | 0.0224 | 0.000548 | **0.0220** | **0.001095** |
| Validation Loss | Best | 0.0242 | 0.001789 | 0.0240 | 0.001732 | 0.0252 | 0.000447 | 0.0247 | 0.001225 |
| Training Dice | Best | 0.8836 | 0.003782 | 0.8800 | 0.006245 | **0.8852** | **0.004324** | 0.8842 | 0.004087 |

| Validation Dice | Best | 0.8714 | 0.006387 | 0.8764 | 0.007570 | 0.8760 | 0.010840 | 0.8740 | 0.008860 |
|---|---|---|---|---|---|---|---|---|---|
| | | | | | | | | | |
| Test on UO | | | | | | | | | |
| Test loss | | 0.19720 | 0.019344 | **0.18100** | **0.063600** | 0.29532 | 0.064662 | 0.29040 | 0.074053 |
| Test Dice | | **0.73254** | **0.016736** | 0.72778 | 0.056359 | 0.65452 | 0.044665 | 0.66786 | 0.049905 |
| MSD (pixels) | | 0.26962 | 0.013043 | 0.26050 | 0.021607 | 0.27034 | 0.033918 | **0.25154** | **0.026763** |
| MSD (mm) | | 0.04040 | 0.001946 | 0.03902 | 0.003274 | 0.04050 | 0.005079 | **0.03768** | **0.004014** |
| | | | | | | | | | |
| Test on SS | | | | | | | | | |
| Test loss | | **0.19530** | **0.026175** | 0.21370 | 0.031392 | 0.22566 | 0.037398 | 0.23354 | 0.059924 |
| Test Dice | | **0.86066** | **0.015557** | 0.83744 | 0.020104 | 0.85592 | 0.022341 | 0.85244 | 0.040555 |
| MSD (pixels) | | 0.38604 | 0.092171 | 0.31790 | 0.072118 | **0.29516** | **0.038401** | 0.32756 | 0.072501 |
| MSD (mm) | | 0.05788 | 0.013795 | 0.04766 | 0.010821 | **0.04422** | **0.005729** | 0.04910 | 0.010879 |

## 4.4 Real-time Performance and Computation Cost

As we discussed before, one goal of this study is to propose a robust, fully-automatic, and at the same time with the capability of real-time performance. Real-time performance of a deep model depends on the network size, hyperparameter tuning, computational facilities, parameter initialization, per- or post data processing, data augmentation method, convergence rate due to the performance of optimization method, activation function and so forth. The first two items are more effective than others on the speed of instantiation of each deep network model as we saw in our experiments. Furthermore, the accuracy of deep learning methods is also highly related to the size of training dataset and the complexity of the deep network model. Hence, there is always a trade-off between the number of training samples, which is a big issue in many applications such as in medicine (Litjens et al., 2017; Ronneberger et al., 2015), and the number of parameters in the network, which it requires more computing and memory units (Badrinarayanan et al., 2015).

Increasing the number of datasets through data acquisition is not always the best and cheapest alternative. Data augmentation can help to alleviate this difficulty, but bigger dataset needs a better network model in terms of generalization. In general, large state of the art deep network models (L.-C. C. Chen et al., 2018; Noh et al., 2015) can be generalized on relatively small datasets such as ultrasound tongue better with the expense of higher computational cost due to the number of parameters. Moreover, training and testing time will rise consequently using big networks. For real-time applications such as ultrasound tongue contour tracking, a desired, robust and accurate network model should also work fast, and smaller network size could be an alternative for this favor.

In this work, we proposed BowNet and wBowNet which are capable of segmenting tongue contour region from ultrasound images automatically with accurate and robust results similar or even better than similar architectures with less parameters. Post-processing stages such as skeletonizing will slightly decrease the performance time while it became considerable when it followed by enhancements. From Table 10, sUNet is the fastest method with double size of the BowNet. It can be seen in the table that the performance speed of each proposed method is in real-time range. Figure 15 shows a sample of training and validating trends for each deep network models in this study on datasets from online augmentation. Although the pattern for all models is similar, slightly fewer fluctuations in train and validation errors can be seen for the wBowNet model. From the figure, one can be seen that sDeepLab is trained slower in the begging than others.

Table 10. A comparison study over each network model in terms the number of parameters, the memory which is required for keeping the parameters of each network models, and frame rate while testing each model on one batch of data. Results are average of 10 times run on each test model with different test datasets and models (using online or offline augmentation).

| | sUNet | sDeepLab | BowNet | wBowNet |
|---|---|---|---|---|
| Number of parameters | 948833 | 785889 | **434785** | 786657 |
| Memory required for learnable parameters | 3795332 | 3143556 | **1739140** | 3146628 |
| Frame-rate using GPU | **72** | 32 | 42 | 30 |

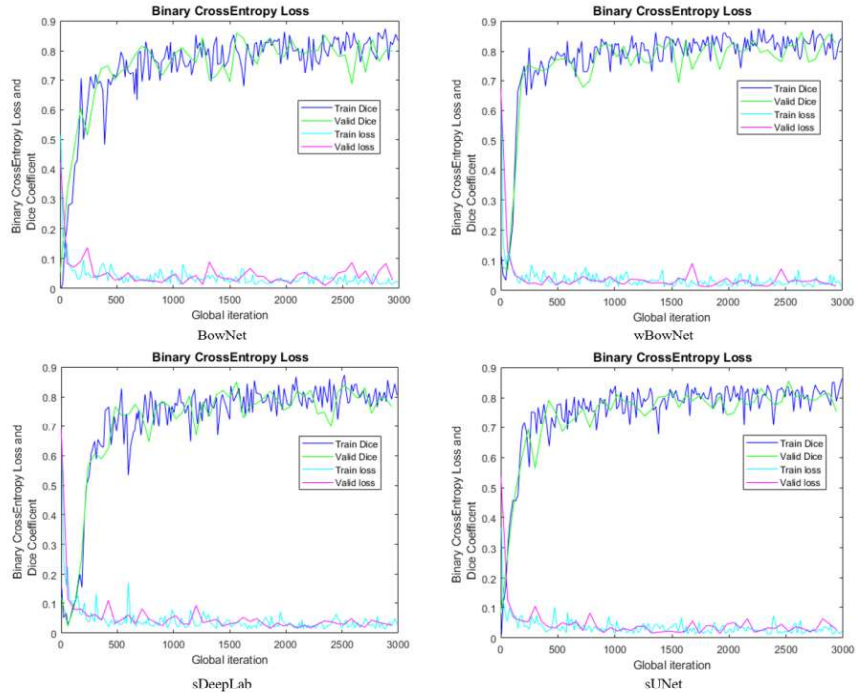

Figure 15. Training and validation trend for the proposed models. Note that for the sake of better presentation, results from every ten iterations is depicted on the horizontal axis.

# 5 Conclusion and Discussion

In this paper, we have proposed and presented two new deep convolutional neural networks called BowNet and wBowNet for tongue contour extraction benefiting from dense image classification and dilated convolution for globally and locally accurate segmentation results. Extensive experimental studies on several types of datasets using online and offline augmentation with rich comparison results demonstrated the outstanding performance of the proposed deep learning techniques. From the evaluation results, it can be concluded that the wBowNet has the most accurate, robust, with the capability of real-time performance using GPU amongst the comparison models such as sUNet, sDeepLab, or even BowNet.

One can conclude that wBowNet is a suitable architecture for segmentation of ultrasound tongue contour images, and it can delineate tongue contours in real-time automatically with utilizing small sized GPU while for more significant GPU powers, we can use UNet like or bigger size BowNet models. On the other hand, sUNet can be the alternative for the cases which CPU only is in demand. The BowNet is a counterpart of the wBowNet with small size network model, can be used for the applications which we need to keep the memory as much as possible while an accurate result is still the desired goal.

Materials of this study might help other researchers in different fields such as linguistics to study tongue gestures in real-time easier, accessible, and with higher accuracy than previous methods. In future work, BowNet models can be applied to other famous databases even non-medical large-scale benchmarks to see the ability of this powerful algorithm. Our qualitative analysis revealed the power of wBowNet architecture in terms of generalizing datasets and handling overfitting issue. The quantitative study can be the support for our observations which it shows that the common comparison criteria for tongue contour segmentation is not suitable anymore whereas architectures with better results in qualitative analysis could not achieve the best results in quantitative evaluations. Finding a robust and accurate approach for comparison of the tongue contour extraction results might be an open direction for study more as a future work. Furthermore, the current infant BowNet models need to be extended for other applications after developing with new assessment criteria.

# References

Abadi, M., Barham, P., Chen, J., Chen, Z., Davis, A., Dean, J., Devin, M., Ghemawat, S., Irving, G., Isard, M., Kudlur, M., Levenberg, J., Monga, R., Moore, S., Murray, D.G., Steiner, B., Tucker, P., Vasudevan, V., Warden, P., Wicke, M., Yu, Y., Zheng, X., 2016. TensorFlow: A system for large-scale machine learning. Methods Enzymol. 101, 582–598. https://doi.org/10.1016/0076-6879(83)01039-3

Akgul, Y.S., Kambhamettu, C., Stone, M., 1999. Automatic extraction and tracking of the tongue contours. IEEE Trans. Med. Imaging 18, 1035–1045. https://doi.org/10.1109/42.811315

Badrinarayanan, V., Kendall, A., Cipolla, R., 2015. SegNet: A Deep Convolutional Encoder-Decoder Architecture for Image Segmentation 1–14. https://doi.org/10.1109/TPAMI.2016.2644615

Bergstra, James, Bengio, Y., 2012. Random Search for Hyper-Parameter Optimization, Journal of Machine Learning Research. https://doi.org/10.1162/153244303322533223

Berry, J., Fasel, I., 2011. Dynamics of tongue gestures extracted automatically from ultrasound, in: Acoustics, Speech and Signal Processing (ICASSP), 2011 IEEE International Conference On. pp. 557–560.

Bloice, M.D., Stocker, C., Holzinger, A., 2017. Augmentor: An Image Augmentation Library for Machine Learning. https://doi.org/10.21105/joss.00432

Chen, L.-C., Papandreou, G., Kokkinos, I., Murphy, K., Yuille, A.L., 2014a. Semantic Image Segmentation with Deep Convolutional Nets and Fully Connected CRFs, Atrous Convolution, and Fully Connected CRFs. Iclr 1–14. https://doi.org/10.1684/ejd.2011.1428

Chen, L.-C., Papandreou, G., Kokkinos, I., Murphy, K., Yuille, A.L., 2014b. Semantic Image Segmentation with Deep Convolutional Nets and Fully Connected CRFs, Atrous Convolution, and Fully Connected CRFs. Iclr 1–14. https://doi.org/10.1684/ejd.2011.1428

Chen, L.-C., Papandreou, G., Schroff, F., Adam, H., 2017. Rethinking Atrous Convolution for Semantic Image Segmentation.

Chen, L.-C.C., Papandreou, G., Kokkinos, I., Murphy, K., Yuille, A.L., 2018. DeepLab: Semantic Image Segmentation with Deep Convolutional Nets, Atrous Convolution, and Fully Connected CRFs. IEEE Trans. Pattern Anal. Mach. Intell. 40, 834–848. https://doi.org/10.1109/TPAMI.2017.2699184

Chen, S., Zheng, Y., Wu, C., Sheng, G., Roussel, P., Denby, B., 2018. Direct, Near Real Time Animation of a 3D Tongue Model Using Non-Invasive Ultrasound Images, in: 2018 IEEE International Conference on Acoustics, Speech and Signal Processing (ICASSP). IEEE, pp. 4994–4998. https://doi.org/10.1109/ICASSP.2018.8462096

Chollet, F., others, 2015. Keras: Deep learning library for theano and tensorflow. URL https//keras. io/k 7, 8.

Denby, B., Schultz, T., Honda, K., Hueber, T., Gilbert, J.M., Brumberg, J.S., 2010. Silent speech interfaces. Speech Commun. 52, 270–287.

Eshky, A., Ribeiro, M.S., Cleland, J., Richmond, K., Roxburgh, Z., Scobbie, J., Wrench, A., 2018. UltraSuite: A Repository of Ultrasound and Acoustic Data from Child Speech Therapy Sessions. Interspeech 1888–1892. https://doi.org/10.21437/Interspeech.2018-1736

Fabre, D., Hueber, T., Bocquelet, F., Badin, P., 2015. Tongue tracking in ultrasound images using eigentongue decomposition and artificial neural networks. Proc. Annu. Conf. Int. Speech Commun. Assoc. INTERSPEECH 2015–Janua, 2410–2414.

Fasel, I., Berry, J., 2010. Deep belief networks for real-time extraction of tongue contours from ultrasound during speech, in: Pattern Recognition (ICPR), 2010 20th International Conference On. pp. 1493–1496. https://doi.org/10.1109/ICPR.2010.369

Ghrenassia, S., Ménard, L., Laporte, C., 2014. Interactive segmentation of tongue contours in ultrasound video sequences using quality maps, in: Medical Imaging 2014: Image Processing. p. 903440.

Gick, B., Bernhardt, B.M., Bacsfalvi, P., Wilson, I., 2008. Ultrasound imging applications in second language acquisition. Phonol. Second Lang. Acquis. 309–322. https://doi.org/10.1684/abc.2012.0768

Hahn-Powell, G. V., Archangeli, D., Berry, J., Fasel, I., 2014. AutoTrace: An automatic system for tracing tongue contours. J. Acoust. Soc. Am. 136, 2104–2104. https://doi.org/10.1121/1.4899570

Hamaguchi, R., Fujita, A., Nemoto, K., Imaizumi, T., Hikosaka, S., 2018. Effective Use of Dilated Convolutions for Segmenting Small Object Instances in Remote Sensing Imagery. Proc. - 2018 IEEE Winter Conf. Appl. Comput. Vision, WACV 2018 2018–Janua, 1442–1450. https://doi.org/10.1109/WACV.2018.00162

Jaumard-Hakoun, A., Xu, K., Roussel-Ragot, P., Dreyfus, G., Denby, B., 2016. Tongue contour extraction from ultrasound images based on deep neural network. Proc. 18th Int. Congr. Phonetic Sci. (ICPhS 2015).

Jaumard-Hakoun, A., Xu, K., Roussel-ragot, P., Stone, M.L., 2015. Tongue Contour Extraction From Ultrasound Images. Proc. 18th Int. Congr. Phonetic Sci. (ICPhS 2015).

Kingma, D.P., Ba, J., 2014. Adam: A method for stochastic optimization. arXiv Prepr. arXiv1412.6980.

Krizhevsky, A., Sutskever, I., Hinton, G.E., 2017. ImageNet Classification with Deep Convolutional Neural Networks. Commun. ACM 60. https://doi.org/10.1145/3065386

L., T., T., B., G., H., Tang, L., Bressmann, T., Hamarneh, G., 2012. Tongue contour tracking in dynamic ultrasound via higher-order MRFs and efficient fusion moves. Med. Image Anal. 16, 1503–1520. https://doi.org/10.1016/j.media.2012.07.001

Laporte, C., Ménard, L., 2018. Multi-hypothesis tracking of the tongue surface in ultrasound video recordings of normal and impaired speech. Med. Image Anal. 44, 98–114. https://doi.org/10.1016/j.media.2017.12.003

Laporte, C., Ménard, L., 2015. Robust tongue tracking in ultrasound images: a multi-hypothesis approach, in: Sixteenth Annual Conference of the International Speech Communication Association.

Lawson, E., Stuart-Smith, J., Scobbie, J.M., Nakai, S., Beavan, D., Edmonds, F., Edmonds, I., Turk, A., Timmins, C., Beck, J.M., others, Esling, J., LePlatre, G., Cowen, S., Barras, W., Durham, M., 2015. Seeing Speech: an articulatory web resource for the study of phonetics [website].

Li, K., Hariharan, B., Malik, J., 2015. Iterative Instance Segmentation. https://doi.org/10.1109/CVPR.2016.398

Li, M., Kambhamettu, C., Stone, M., 2005. Automatic contour tracking in ultrasound images. Clin. Linguist. Phonetics 19, 545–554. https://doi.org/10.1080/02699200500113616

Lin, G., Milan, A., Shen, C., Reid, I., 2017. RefineNet: Multi-path refinement networks for high-resolution semantic segmentation, Proceedings - 30th IEEE Conference on Computer Vision and Pattern Recognition, CVPR 2017. https://doi.org/10.1109/CVPR.2017.549

Litjens, G., Kooi, T., Bejnordi, B.E., Setio, A.A.A., Ciompi, F., Ghafoorian, M., van der Laak, J.A.W.M.W.M., van Ginneken, B., Sánchez, C.I., 2017. A survey on deep learning in medical image analysis. Med. Image Anal. 42, 60–88. https://doi.org/10.1016/j.media.2017.07.005

Liu, X.Y., Wu, J., Zhou, Z.H., 2006. Exploratory under-sampling for class-imbalance learning, in: Proceedings - IEEE International Conference on Data Mining, ICDM. pp. 965–969. https://doi.org/10.1109/ICDM.2006.68

Long, J., Shelhamer, E., Darrell, T., 2015. Fully convolutional networks for semantic segmentation, in: Proceedings of the IEEE Conference on Computer Vision and Pattern Recognition. pp. 3431–3440.

Martínková, N., Nová, P., Sablina, O. V., Graphodatsky, A.S., Zima, J., 2004. Karyotypic relationships of the Tatra vole (Microtus tatricus). Folia Zool. 53, 279–284. https://doi.org/10.1007/s13398-014-0173-7.2

Milletari, F., Navab, N., Ahmadi, S.A., 2016. V-Net: Fully convolutional neural networks for volumetric medical image segmentation, in: Proceedings - 2016 4th International Conference on 3D Vision, 3DV 2016. IEEE, pp. 565–571. https://doi.org/10.1109/3DV.2016.79

Mozaffari, M.H., Guan, S., Wen, S., Wang, N., Lee, W.-S., 2018. Guided Learning of Pronunciation by Visualizing Tongue Articulation in Ultrasound Image Sequences, in: 2018 IEEE International Conference on Computational Intelligence and Virtual Environments for Measurement Systems and Applications (CIVEMSA). IEEE, pp. 1–5. https://doi.org/10.1109/CIVEMSA.2018.8440000

Noh, H., Hong, S., Han, B., 2015. Learning Deconvolution Network for Semantic Segmentation. Proc. IEEE Int. Conf. Comput. Vis. 2015 Inter, 1520–1528. https://doi.org/10.1109/ICCV.2015.178

Odena, A., Dumoulin, V., Olah, C., 2016. Deconvolution and Checkerboard Artifacts. Distill 1, e3. https://doi.org/10.23915/distill.00003

Ohkubo, M., Scobbie, J.M., 2018. Tongue Shape Dynamics in Swallowing Using Sagittal Ultrasound. Dysphagia 1–7. https://doi.org/10.1007/s00455-018-9921-8

Ronneberger, O., Fischer, P., Brox, T., 2015. U-net: Convolutional networks for biomedical image segmentation, in: International Conference on Medical Image Computing and Computer-Assisted Intervention. pp. 234–241.

Simonyan, K., Zisserman, A., 2014. Very Deep Convolutional Networks for Large-Scale Image Recognition.

Stone, M., 2005. A guide to analysing tongue motion from ultrasound images. Clin. Linguist. Phonetics 19, 455–501. https://doi.org/10.1080/02699200500113558

Szegedy, C., Liu, W., Jia, Y., Sermanet, P., Reed, S., Anguelov, D., Erhan, D., Vanhoucke, V., Rabinovich, A., 2014. Going Deeper with Convolutions. Popul. Health Manag. 18, 186–191. https://doi.org/10.1089/pop.2014.0089

Tang, L., Hamarneh, G., 2010. Graph-based tracking of the tongue contour in ultrasound sequences with adaptive temporal regularization. 2010 IEEE Comput. Soc. Conf. Comput. Vis. Pattern Recognit. - Work. CVPRW 2010 154–161. https://doi.org/10.1109/CVPRW.2010.5543597

Thoma, M., 2016. A Survey of Semantic Segmentation.

Xu, K., Gábor Csapó, T., Roussel, P., Denby, B., 2016a. A comparative study on the contour tracking algorithms in ultrasound tongue images with automatic re-initialization. J. Acoust. Soc. Am. 139, EL154-EL160. https://doi.org/10.1121/1.4951024

Xu, K., Yang, Y., Jaumard-Hakoun, A., Leboullenger, C., Dreyfus, G., Roussel, P., Stone, M., Denby, B., 2016b. Development of a 3D tongue motion visualization platform based on ultrasound image sequences. arXiv Prepr. arXiv1605.06106.

Xu, K., Yang, Y., Stone, M., Jaumard-Hakoun, A., Leboullenger, C., Dreyfus, G., Roussel, P., Denby, B., 2016c. Robust contour tracking in ultrasound tongue image sequences. Clin. Linguist. Phon. 30, 313–327. https://doi.org/10.3109/02699206.2015.1110714

Yu, F., Koltun, V., 2015. Multi-Scale Context Aggregation by Dilated Convolutions.

Zeiler, M.D., Fergus, R., 2014. Visualizing and understanding convolutional networks, in: Lecture Notes in Computer Science (Including Subseries Lecture Notes in Artificial Intelligence and Lecture Notes in Bioinformatics). Springer, Cham, pp. 818–833. https://doi.org/10.1007/978-3-319-10590-1_53

Zeiler, M.D., Krishnan, D., Taylor, G.W., Fergus, R., 2010. Deconvolutional networks, in: 2010 IEEE Computer Society Conference on Computer Vision and Pattern Recognition. IEEE, pp. 2528–2535. https://doi.org/10.1109/CVPR.2010.5539957

Zhu, J., Styler, W., Calloway, I.C., 2018. Automatic tongue contour extraction in ultrasound images with convolutional neural networks. J. Acoust. Soc. Am. 143, 1966–1966. https://doi.org/10.1121/1.5036466